\title{Comprehensive evaluation of differential expression analysis methods for RNA-seq data}
\author{
 Franck Rapaport $^1$, Raya Khanin $^1$, Yupu Liang $^1$, Azra Krek $^1$, Paul Zumbo $^{2,4}$, \\
 Christopher E. Mason $^{2,4}$, Nicholas D. Socci $^1$, {Doron Betel} $^{3,4}$\\
\\
$^{1}$\small{Bioinformatics Core, Memorial Sloan-Kettering Cancer Center, New York}\\
$^{2}$\small{Department of Physiology and Biophysics, Weill Cornell Medical College, New York}\\
$^{3}$ \small{Division of Hematology/Oncology, Department of Medicine, Weill Cornell Medical College, New York}\\
$^{4}$ \small{Institute for Computational Biomedicine, Weill Cornell Medical College, New York}}

\date{\today}

\documentclass[12pt]{article}
\usepackage{fullpage}
\usepackage{hyperref}
\usepackage{graphicx}
\usepackage{url}
\usepackage{soul}
\usepackage{caption}
\usepackage{subcaption} 
\usepackage{lscape} 
\usepackage{floatpag} 
\usepackage{float}
\begin{document}
\maketitle

\newcommand{\logTwo}{$\log_2$}

\begin{abstract}
High-throughput sequencing of RNA transcripts (RNA-seq) has become the method of choice for detection of differential expression (DE). Concurrent with the growing popularity of this technology there has been a significant research effort devoted towards understanding the statistical properties of this data and developing analysis methods. We report on a comprehensive evaluation of the commonly used DE methods using the SEQC benchmark data set. We evaluate a number of key features including : assessment of normalization, accuracy of DE detection, modeling of genes expressed in only one condition, and the impact of sequencing depth and number of replications on identifying DE genes. We find significant differences among the methods with no single method consistently outperforming the others. Furthermore, the performance of array-based approach is comparable to methods customized for RNA-seq data. Perhaps most importantly, our results demonstrate that increasing the number of replicate samples provides significantly more detection power than increased sequencing depth. 
\end{abstract}

\section*{Introduction}
High-throughput sequencing technology is rapidly becoming the standard method for measuring RNA expression levels (a.k.a. RNA-seq) \cite{Mortazavi:2008zr}. The advent of rapid sequencing technologies along with reduced costs has enabled detailed profiling of gene expression levels, impacting almost every field in life sciences and is now being adopted for clinical use \cite{Berger:2010fk}. RNA-seq technology enables the detailed identification of gene isoforms, translocation events, nucleotide variations and post transcriptional base modifications \cite{Wang:2009kx}. One of the main goals of these experiments is to identify the differentially expressed genes between two or more conditions. Such genes are selected based on a combination of expression change threshold and score cutoff, which is usually based on p-values generated by statistical modeling. 

The expression level of each RNA unit is measured by the number of sequenced fragments that map to the transcript, which is expected to correlate directly with its abundance level. This measure is fundamentally different from gene probe-based methods, such as microarrays : in RNA-seq the expression signal of a transcript is limited by the sequencing depth and is dependent on the expression levels of other transcripts, whereas in array-based methods probe intensities are independent of each other. This, as well as other technical differences, has motivated a growing number of statistical algorithms that implement a variety of approaches for normalization and differential expression (DE) detection. 
Typical approaches use Poisson or Negative Binomial distributions to model the gene count data and a variety of normalization procedures (see \cite{Young:2012} for additional review). 

In this comparison study we evaluated the more commonly used and freely available differential expression software packages: Cuffdiff\cite{Trapnell:2010ve}, edgeR\cite{Robinson:2010fk}, DESeq\cite{Anders:2010uq}, PoissonSeq\cite{Li:2012ly}, baySeq\cite{Hardcastle:2010fk}, and limma\cite{Smyth:2004nx}  adapted for RNA-seq use. We used the standardized Sequencing Quality Control (SEQC) datasets which include replicated samples of the human whole body reference RNA and human brain reference RNA along with RNA spike-in controls. These samples are part of the MAQC study for benchmarking microarray technology \cite{Shi:2010vn} as well as the SEQC effort to characterize RNA-seq technology and include  close to 1,000 genes that were validated by TaqMan qPCR. Our evaluations are focused on a number of measures that are most relevant for detection of differential gene expression from RNA-seq data: i) normalization of count data; ii) sensitivity and specificity of DE detection; iii) performance on the subset of genes that are expressed in one conditions but have no detectable expression in the other condition; iv) and  finally, we investigated the effects of reduced coverage and number of replicates on the detection of differential expression.

Our results demonstrate substantial differences among the methods both in terms of specificity and sensitivity of the detection of differential expressed genes. In most benchmarks Cuffdiff performed less favorably with a higher number of false positives without any increase in sensitivity. Our results conclusively demonstrate that the addition of replicate samples provide substantially greater detection power of DE than increased sequence coverage. Hence, including more replicate samples in RNA-seq experiments is always preferred over increased coverage.

\subsection*{Theoretical Background}
A convenient starting point for comparing different RNA-seq analysis methods is to begin with a simple count matrix $\mathbf{N}$ of $n \times m$ where $N_{ij}$ is the number of reads assigned to (i.e. coverage of) gene $i$ in sequencing experiment $j$. Such matrices can be produced from alignment data using tools such as HTSeq \cite{HTseq}, picard \cite{picard}, BEDTools \cite{Quinlan:2010fk} or Cuffdiff \cite{Trapnell:2010ve}. The work presented here does not address the important subtleties when calculating gene counts, in particular which gene model to use and the use of ambiguously mapped reads. Rather, the focus is on the comparison between methods given a fixed expression count matrix. For Cuffdiff, which uses a different quantitation method that is not compatible with the others, we used its joined method Cufflinks and for all other methods we used HTSeq.  

Differential gene expression analysis of RNA-seq data generally consists of three components: normalization of counts, parameter estimation of the statistical model and tests for differential expression. In this section we provide a brief background into the approaches implemented by the various algorithms that perform these three steps. We limit our discussion to the most common case of measuring differential expression between two cellular conditions or phenotypes although some of the packages can test for multi-class differences or multi-factored experiments where multiple biological conditions and different sequencing protocols are included.

\subsubsection*{Normalization} 
The first difficulty to address when working with sequencing data is the large differences in number of reads produced between different sequencing runs as well as technical biases introduced by library preparation protocols, sequencing platforms and nucleotide compositions \cite{Dillies:2012fk}. Normalization procedures attempt to account for such differences in order to facilitate accurate comparisons between sample groups. An intuitive normalization is to simply divide the gene count by the total number of reads in each library, or mapped reads, as first introduced by Motazavi et al. \cite{Mortazavi:2008zr}, a normalization procedure named Reads per Kilobase per Million reads (RPKM). A deficiency of this approach is that the proportional representation of each gene is dependent on the expression levels of all other genes. Often a small fraction of genes account for large proportions of the sequenced reads and small expression changes in these highly expressed genes will skew the counts of lowly expressed genes under this scheme. This can result in erroneous differential expression \cite{Bullard:2010ys,Robinson:2010kl}.  A variation of the RPKM termed fragments per kilobase of exon per million mapped fragments (FPKM) was introduced by Trapnell et al. to accommodate paired-end reads \cite{Trapnell:2010ve}, however this has the same limitation of coupling changes in expression levels among all genes. DESeq computes a scaling factor for a given sample by computing the median of the ratio, for each gene, of its read count over its geometric mean across all samples. It then uses the assumption that most genes are not DE and uses this median of ratios to obtain the scaling factor associated with this sample. Cuffdiff extends this by first performing intra-condition library scaling and then a second scaling between conditions. Cuffdiff attempts to also explicitly account for changes in isoform levels by additional transcript-specific normalization that estimates the abundance of each isoform.   

Other normalization procedures attempt to use a subset of stably expressed genes or to normalize within replicated samples to globally adjust library sizes. The Trimmed Means of M values (TMM) from Robinson et al. \cite{Robinson:2010kl}, which is implemented in edgeR, computes a scaling factor between two experiments by using the weighted average of subset of genes after excluding genes that exhibit high average coverage and genes that have large differences in expression. Another approach is to sum gene counts up to upper 25\% quantile to normalize library sizes as proposed by Bullard et al. \cite{Bullard:2010ys} and is the default normalization in the baySeq package. 
The PoissonSeq package uses a goodness-of-fit estimate to define a gene set that is least differentiated between two conditions which is then used to compute library normalization factors. Quantile normalization ensures that the counts across all samples have the same empirical distribution by sorting the counts from each sample and setting the values to be equal to the quantile mean from all samples \cite{Bolstad:2003fk}. This normalization is widely used in expression arrays and is implemented in the limma package. Recently, a new normalization function termed \texttt{voom} was added to the limma package designed specifically for RNA-seq data which performs a lowess regression to estimate the mean-variance relationship and transforms the read counts to the appropriate log form for linear modeling.  

\subsubsection*{Statistical Modeling of Gene Expression} 
If sequencing experiments are considered as random sampling of reads from a fixed pool of genes then a natural representation of gene coverage is the Poisson distribution of the form
$f(n,\lambda) = (\lambda^ne^{-\lambda}) /{n!}$
where $n$ is the number of read counts and $\lambda$ is a real number equal to the expected number of reads from transcript fragment. An important property of the Poisson distribution is that the variance is equal to the mean, which equals $\lambda$. However, in reality the variance of gene expression across multiple biological replicates is larger than its mean expression values \cite{Robinson:2007bh, Nagalakshmi:2008dq, Marioni:2008oq}. To address this \textit{overdispersion} problem methods such as edgeR and DESeq use the related negative binomial distribution (NB) where the relationship between the variance $\nu$ and mean $\mu$ is defined as $\nu= \mu + \alpha \mu^2$ where $\alpha$ is the dispersion factor.  Estimation of this factor is one of the fundamental differences between the edgeR and DESeq packages. edgeR estimates $\alpha$ as \textit{weighted} combination of two components. The first is a gene-specific dispersion effect and the second is a common dispersion effect calculated from all genes.  DESeq, on the other hand, breaks the variance estimate to a combination of the Poisson estimate (i.e. the mean expression of the gene) and a second term which models the biological expression variability.  Cuffdiff computes a separate variance model for single-isoform genes and multi-isoforms genes. Single-isoform expression variance is computed similarly to DESeq and multi-isoforms variance is modeled by a mixture model of negative binomials using the beta distribution parameters as mixture weights. baySeq implements a full Bayesian model of negative binomial distributions in which the prior probability parameters are estimated by numerical sampling from the data.
PoissonSeq models the gene counts $N_{i,j}$ as a Poisson variable in which the mean $\mu_{i,j}$ of the distribution is represented by the log-linear relationship $\log{\mu_{ij}} = \log{d_j} + \log{\beta_i}+ \gamma_i y_j$; where $d_j$ represents the normalized library size, $\beta_i$ is the expression level of gene $i$ and $\gamma_i$ is the correlation of gene $i$ with condition $y_j$ (note that in \cite{Li:2012ly} the subscripts $i$ and $j$ are samples and genes respectively). If gene $i$ is not correlated with the sample $j$ class (i.e. there is no significant difference in gene $i$ expression between two conditions) then $\gamma_i$ is zero.  

\subsubsection*{Test for Differential Expression}
The estimation of the parameters for the respective statistical model is followed by the test for differential expression, the calculation of the significance of change in expression of gene $i$ between two conditions. Both edgeR and DESeq use a variation of the Fisher exact test adopted for NB distribution hence they return exact p-values computed from the derived probabilities. Cuffdiff uses the test statistics T=$E[log(y)]/Var[log(y)]$, where $y$ is the ratio of the normalized counts between two conditions, and this ratio approximately follows a normal distribution. Hence a t-test is used to calculate the p-value for DE. Limma uses a moderated t-statistic to compute p-values in which both the standard error and the degrees of freedom are modified \cite{Smyth:2004nx}. The standard error is moderated across genes with a shrinkage factor which effectively borrows information from all genes to improve the inference on any single gene. The degrees of freedom are also adjusted by a term that represents the a priori number of degrees for freedom for the model. The baySeq approach estimates two models for every gene, one assuming no differential expression and a second assuming differential expression using the two sample groups. The ratio between the posterior probabilities of the two models is the likelihood for DE. In the PoissonSeq method the test for differential expression is simply a test for the significance of the $\gamma_i$ term (i.e. correlation of gene $i$ expression with the two conditions) which is evaluated by score statistics. By simulation experiments it was shown that this score statistics follows chi-squared distribution which is used to derive p-values for DE. All methods use standard approaches for multiple hypothesis correction (e.g. Benjamini-Hochberg) with exception of PoissonSeq, which implemented a novel estimation of FDR that is specific for the Poisson distribution.   

\section*{Methods}\label{Methods}
\subsection*{Datasets}
In this study, we used samples from two biological sources that were part of the SEQC study, each generated from a mixture of biological sources and a set of synthetic RNAs from the External RNA Control Consortium (ERCC) at known concentrations.
The samples from group (\textbf{A}) contain the Strategene Universal Human Reference RNA (UHRR), which is composed of total RNA from 10 human cell lines, with 2\% by volume of ERCC mix 1. The second group of samples (\textbf{B}) contains the Ambion Human Brain Reference RNA (HBRR) with 2\% by volume of ERCC mix 2. The ERCC spike-in control is a mixture of 92 synthetic polyadenylated oligonucleotides of 250-2000 nucleotides long that are meant to resemble human transcripts. The two ERCC mixtures in groups A and B contain different concentrations of four subgroups of the synthetic spike-ins such that the log expression change is predefined and can be used to benchmark DE performance (see methods section in Li et al. in main SEQC submission). Four replicate libraries from groups A and B were prepared by a single technician and a fifth sample was prepared by Illumina for a total of 10 libraries. All libraries were sequenced as paired-end 100 bases in the Epigenomics Core facility at Weill Cornell Medical College with a full block design on two flow cells on a single HiSeq2000 instrument. We note that these samples are considered technical replicates and therefore represent an idealized scenario of minimal variation.   

\subsection*{Sequence Alignment and Gene Counts}
All sequenced libraries were mapped to the human genome (hg19) using TopHat(v.2.0.3)\cite{Trapnell:2009uq} with the following parameters: \texttt{-r 70 --mate-std-dec 90}. A custom GTF file that includes both RefSeq information (from UCSC genome browser) and the ERCC transcript information was used (\texttt{--GTF \$SEQCLB/hg19\_150\_ERCC.gtf}) along with the transcriptome index option (\texttt{--transcriptome-index \$SEQCLIB/hg19\_150\_ERCC}). Genes shorter than 150 bp were excluded from this GTF file.

HTSeq (v.0.5.3p3) \cite{HTseq} was used to generate the count matrix with the following parameters: \texttt{htseq-count -m intersection-strict -s no} with the same  GTF file used for the alignment step (\texttt{\$SEQCLIB/hg19\_150\_ERCC.gtf}).

\subsection*{Normalization and Differential Expression}
With the exception of Cuffdiff, all differential expression analysis was performed using the same gene count matrix output from HTSeq. Analysis followed the procedures and steps described in the package's documentations and unless stated otherwise default parameters were used in all function calls. Adjusted p-values for multiple hypothesis corrections were used as calculated by the methods. 
The following are the details for each package used in this study:
   
\begin{itemize}
\item \texttt{DESeq} (v.1.10.1): dispersion estimate call to \texttt{estimateDispersions} with parameters \texttt{method= "per-condition"} and \texttt{fitType="local"} and for null model evaluation with no replicates \texttt{method= "blind"}, \texttt{fitType="local"}, \texttt{sharingMode="fit-only"}.   
\item \texttt{edgeR} (v. 3.0.2): In null model comparison with no replicates the \texttt{common.dispersion} value was set to 0.4 as suggested by the documentation.  
\item \texttt{PoissonSeq} (v.1.1.2): No minimum expression mean was applied and number of permutations was 500. 
\item \texttt{baySeq} (v.1.12.0): Sequence length correction was added to normalization as suggested in the documentation. Negative binomial parameter estimation was performed using \texttt{getPriors.NB} using quasi-likelihood estimation. Note that \texttt{baySeq} reports posterior probabilities for differences between two models and not p-values.
\item \texttt{limma(v.3.14.1)} Analysis was performed in two modes which differ in the normalization procedure. Quantile normalization was performed on the \logTwo\ transformed gene counts (with addition of 1 to avoid log of 0) by \texttt{normalizeBetweenArrays} function (henceforth \texttt{limmaQN}). In the second mode, counts were normalized using the \texttt{voom} function where library sizes were scaled by edgeR normalization factors and the mean-variance trend was calculated using lowess regression (henceforth \texttt{limmaVoom}). Note that \texttt{limma} does not allow contrasting libraries with no replication and therefore \texttt{limma} was excluded from the single library comparisons.
\item \texttt{cuffdiff (v.2.0.0 (3365))} with the options: \texttt{--no-update-check --emit-count-tables}, GTF file \texttt{\$SEQCLIB/hg19\_150\_ERCC.gtf}.
\end{itemize}

For each method, comparisons were performed between the five replicates from sample type \textbf{A} with the five replicates from type \textbf{B}. In the null model comparison two models were tested, with replication and without replication. In the replication model, replicates from the same samples were contrasted: \{A1,A2\} vs. \{A3 A4\}, \{A1,A2\} vs. \{A3, A4, A5\}, and \{B1,B2\} vs. \{B3, B4\}. Comparisons without replication were performed between the following samples: A1 vs. A2, A3 vs. A4, B1 vs. B2, B3 vs. B4.

\subsection*{Sample Clustering}
Normalized counts were \logTwo\ transformed after addition of pseudo counts. For counts produced by HTSeq the pseudo counts were set to the smallest non-zero gene count in each library and for FPKM data the pseudo count was set to 0.001. Clustering was performed using the R \texttt{hclust} function with the Euclidean distance measure.

\subsection*{Random Sampling and Coverage Depth}
To assess the effect of a reduced coverage depth, we used \texttt{DownsampleSam}, function from Picard \cite{picard} that randomly samples read pairs, from a SAM file, using a uniform probability. We generated a first set of reduced coverage depth sample by downsampling every file A1 to A5 and B1 to B5 with a probability of $p_1 = 0.5$ for retaining each read. We then downsampled the resulting files with a probability $p_2 = 0.8$ therefore, generating a set that downsampled the original files with a probability $p_1 \times p_2 = 0.4$ representing 40\% sequencing depth. We continued this downsampling cascade, ultimately generating six sets of files with respective probabilities $0.5$, $0.4$, $0.3$, $0.2$, $0.1$ and $0.05$ of retaining each pair of reads from the original files. We then repeated the operation five times, generating five random data sets for each probability value.

For each downsampling probability, we looked at five independent samplings and computed differential expression analysis for every combination of two samples from the A set and two samples from the B set, three samples from each set, four samples from each set as well as the whole A and B sets. We evaluated the DE using DESeq, edgeR, PoissonSeq, and limma using the two described modes.

\paragraph{Source Code:} Source code and data files are available at  \url{http://bitbucket.org/soccin/seqc}.

\section*{Results}\label{Results}
\subsection*{Assessment of normalized counts by sample clustering and log expression correlation}
Normalization of read counts is a critical step in the analysis of RNA-seq data. It attempts to control for the differences in sequencing depths such that gene expression levels can be directly comparable across different samples. In addition, some normalization methods can be used to correct for other effects such as variations in GC content and transcript length \cite{Dillies:2012fk}. To evaluate the normalization approaches we performed hierarchical clustering of samples after \logTwo\ transformation of the normalized count values. Overall, all methods achieved perfect separation between group A and B samples (Supplementary Figure 1). 
Dunn cluster validity index, which measures the ratios of \textit{inter}-cluster over \textit{intra}-cluster distances, indicates that edgeR and Cuffdiff have reduced cluster separation (avg. Dunn index 2.58) relative to the other methods (avg. Dunn index 3.75, Supplementary Figure 2). $Log_2$ distribution of the normalized read counts are similar among most methods with the exception of limmaVoom, and Cuffdiff (Supplementary Figure 3) presumably due to the gene-specific normalization approaches by those two methods in contrast to the global scaling that is used by the other methods.   

\begin{figure}[!ht]
\includegraphics[scale=0.65]{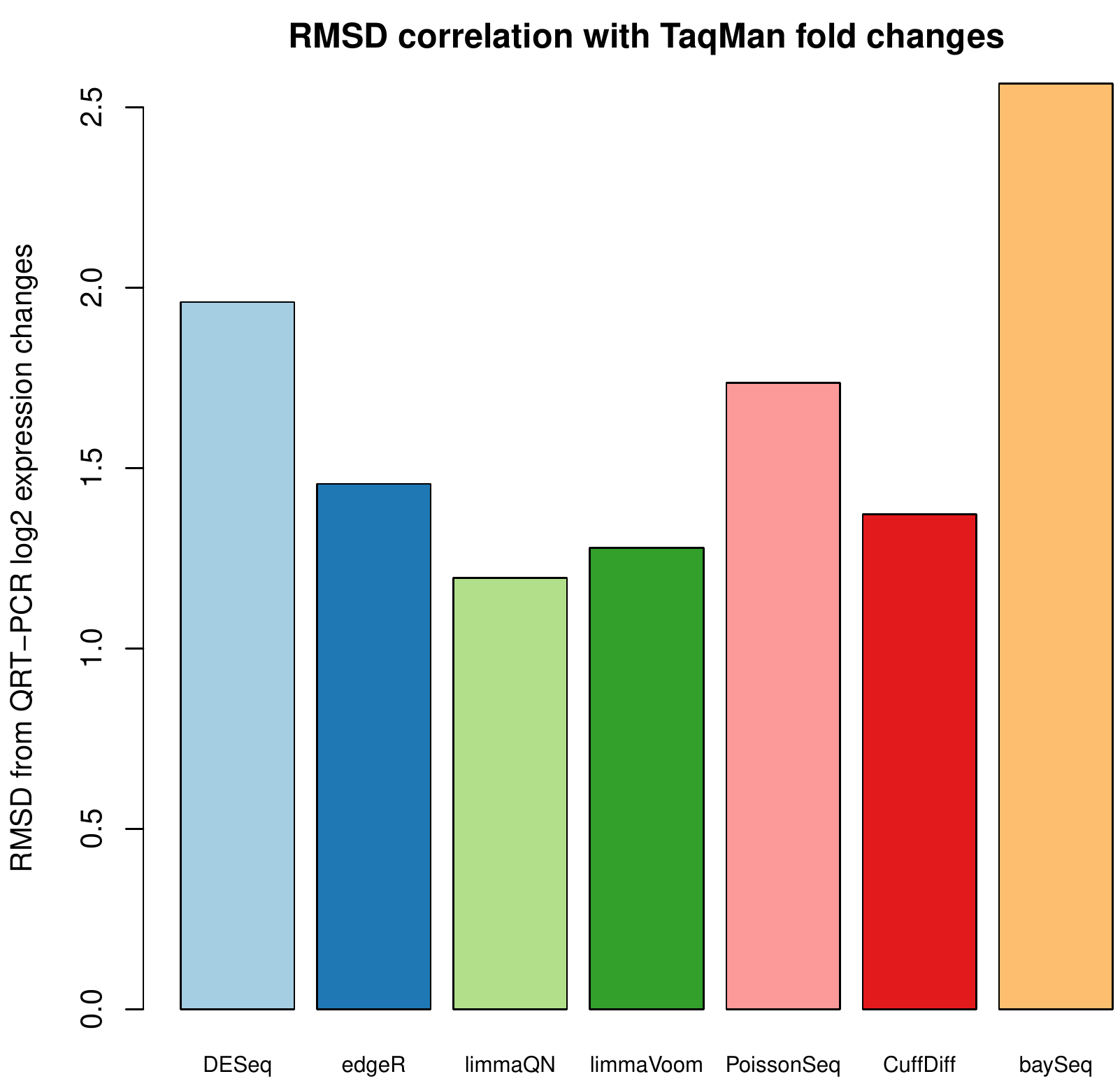}
\caption{\footnotesize{\textbf{RMSD correlation between QRT-PCR and RNA-seq \logTwo\ expression changes computed by each method}. Upper quartile normalization implemented in baySeq package is least correlated with QRT-PCR values.}}
\label{rmsd}
\end{figure}

As an additional measure of the accuracy of normalization we correlated the \logTwo\  normalized expression changes reported by each method with log expression changes measured by QRT-PCR \cite{Canales:2006fk}. Since expression changes are unit-less measures (i.e. this is a ratio of two expression values) we expect the changes to be similar in magnitude and in range regardless of the measurement platform. To assess how accurately the methods matched the PCR data, we used root-mean-square deviation (RMSD) to measure the \textit{difference} of the reported expression changes to the PCR standard. We found that all methods performed well with average RMSD accuracy of 1.65 (and Pearson correlation of 0.92) (Figure \ref{rmsd}). 

\subsection*{Differential expression analysis}
We next evaluated the ability of the various methods to detect differentially expressed genes using both the ERCC and TaqMan data. The ERCC data contains a mixture of spike-in synthetic oligonucleotides that are mixed into samples A and B at four mixing ratios of 1/2, 2/3, 1 and 4. It is, therefore, possible to test how well the methods correctly identify these ratios. Using the mixing ratios of 1:1 ( log ratio = 0) as the true negative set and all others as true positive we performed an ROC analysis to compare the performance of the various methods in detecting differentially mixed spike-in controls. Overall, all methods performed reasonably well in detecting the truly differentiated spiked-in sequences with an average area under the curve (AUC) of 0.78 (Supplementary Figure 4). 

\begin{figure}[htp]
\begin{center}
\begin{tabular}{ll}
 {\includegraphics[scale=.48]{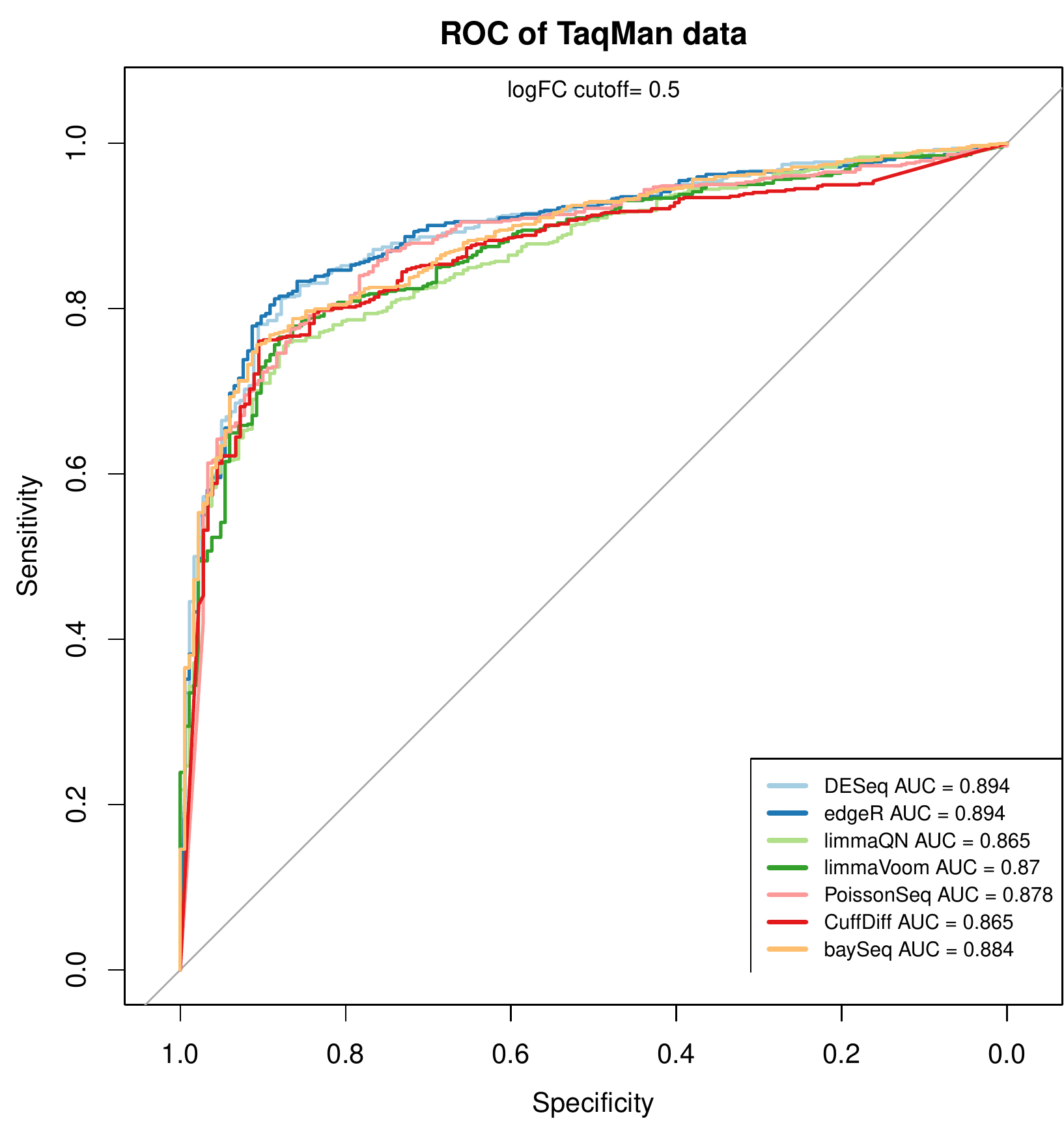}  }  & {\includegraphics[scale=.52]{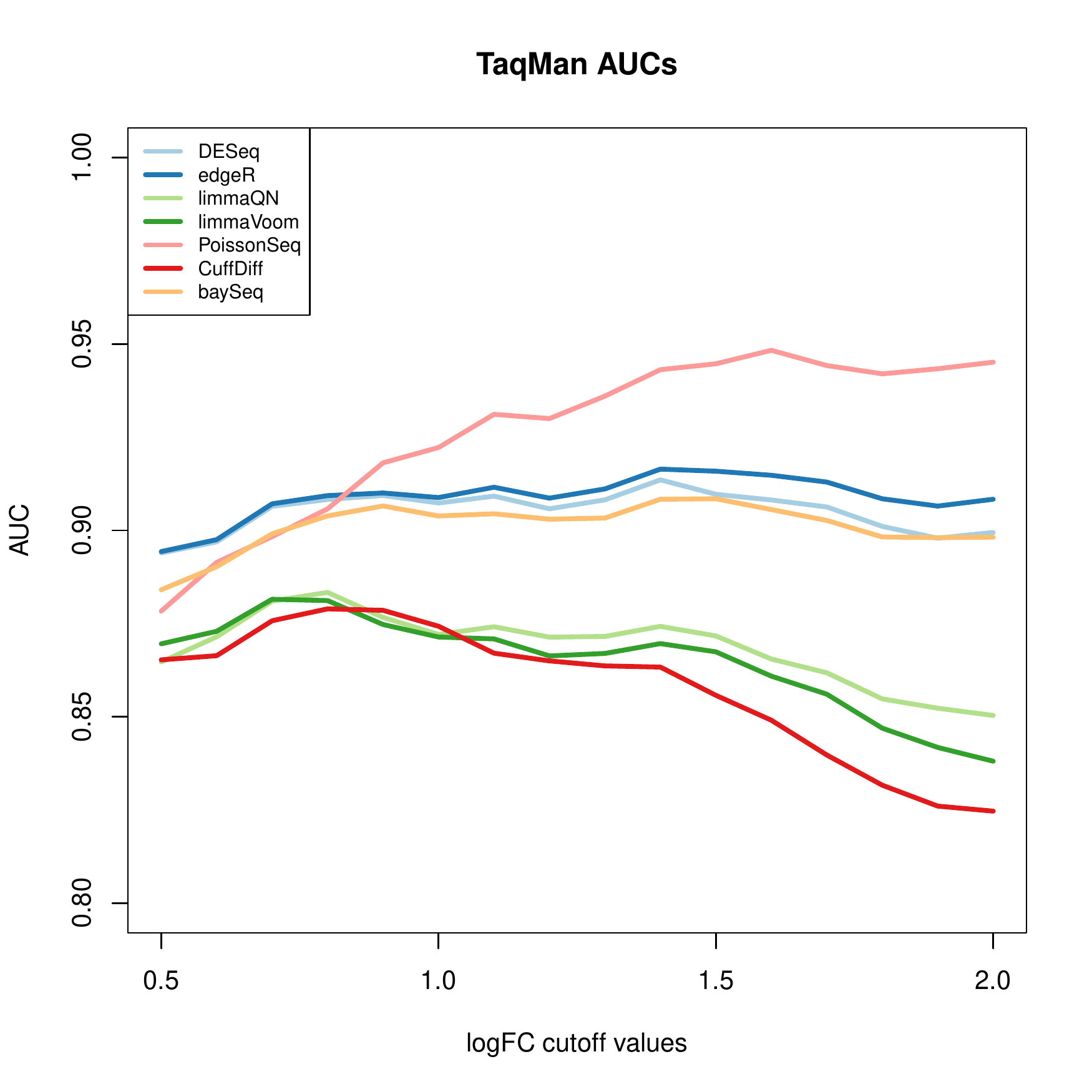}  } \\
 {(a)} & {(b)} \\
\end{tabular}
\caption{\footnotesize{\textbf{Differential expression analysis using QRT-PCR validated gene set.} ({\bf a}) ROC analysis at \logTwo\ expression change cutoff of 0.5 indicates a slight advantage for DESeq and edgeR in detection accuracy. ({\bf b}) At increasing \logTwo\ expression ratio thresholds (incremented by 0.1), representing more stringent cutoff for differential expression, the performances of Cuffdiff and limma methods gradually reduce whereas PoissonSeq performance increases. }  \label{TaqMan}
}
\end{center} 
\end{figure}

A more comprehensive control group is the set of roughly 1000 genes whose expression changes were previously measured by QRT-PCR that span a wider range of expression ratios and represent a sampling of the human transcripts \cite{Canales:2006fk}. We performed a ROC analysis using \logTwo\  expression change cutoff of 0.5 (1.4x expression change measured by QRT-PCR) as the threshold for true differentiation. The AUC values at this cutoff indicate comparable performance among all methods with a slight advantage for DESeq and edgeR (Figure \ref{TaqMan}a). We extended this analysis by measuring AUC at increasing values of the cutoff, which define sets of differentially expressed genes at increasing stringency (Figure \ref{TaqMan}b). Here we find a significant performance advantage for negative binomial and Poisson-based approaches with consistent AUC values at $\sim$0.9 or higher in contrast to Cuffdiff and limma methods with decreasing AUCs indicating reduced discrimination power at higher log fold change values.     

\subsection*{Null model evaluation of Type-I errors}
\begin{figure}[h!]
\includegraphics[scale=0.75]{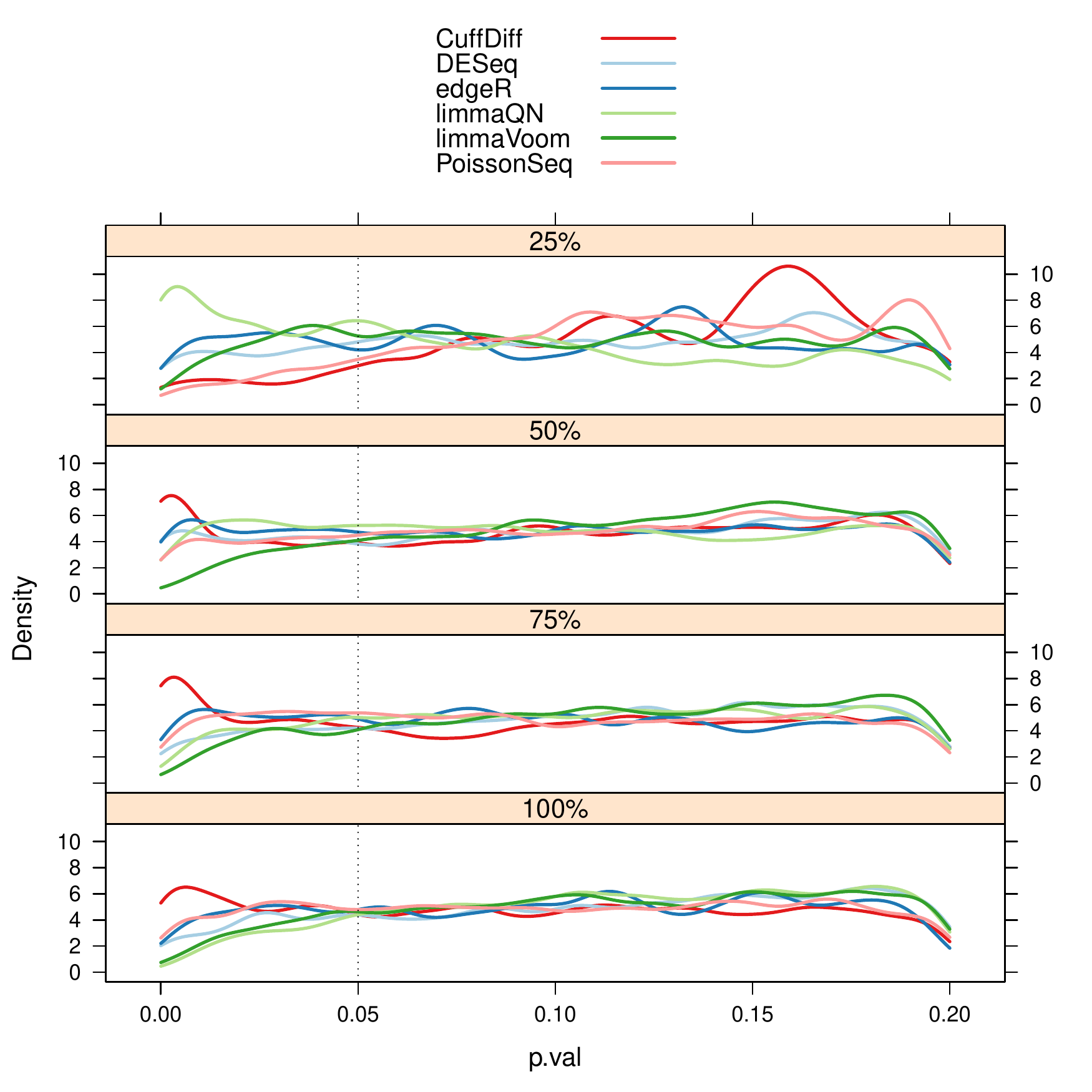}
\caption{\footnotesize{\textbf{Density plots of p-value distributions of null model comparisons by gene expression quantiles.} At the significance range of $\leq$ 0.05 there is a noticeable increase in p-value densities in Cuffdiff results. Smoothing bandwidth was fixed at 0.0065 for all density plots and 25\% is the lowest expression quartile.}}
\label{nullModel}
\end{figure}

A primary goal for any differential expression algorithm is to minimize Type-I errors, which are incorrect rejections of the null hypothesis $H_{o}: \mu_{i,A} = \mu_{i,B}$, where $\mu_{i,A/B}$ is the mean expression of gene $i$ in condition A or B, resulting in a false prediction of differential expression (false positive). To test the number of false positive predictions from the null models we performed a series of intra-condition comparisons using the replicate samples from each condition (see Methods). No genes are expected to be differentially expressed in these comparisons and the distribution of p-values is expected to be uniform since they are derived from the null model. We note that baySeq was excluded from this analysis since it reports posterior probabilities of a model and not p-values, which does not allow us to control it with the same stringency as other methods. We indeed found that the p-values for all methods were largely uniform although less so for the lower 25\% expressed genes where experimental noise is larger than the expression signal (Figure \ref{nullModel}). A noticeable exception was increased number of p-values at the lower range ($\leq 0.05$) for Cuffdiff distribution indicating large number of false positives. A similar observation was noted by Anders et al. where Cuffdiff had inflated number of false positive predictions in their null model comparison \cite{Anders:2012fk}. This trend was even more pronounced when the null model comparison was performed without replicated samples (e.g. Sample A\_1 vs. Sample A\_2, Supplementary Figure 5). Table 1 summarizes the number of false-positive predictions identified by each method at \textit{adjusted} p-values cutoff (or FDR) of $\leq 0.05$. Although the number of false predictions is below the 5\% false discovery rates the reduced specificity points to inflation of differential expression detection by Cuffdiff.  When the comparison is performed with no replicated samples Cuffdiff false discovery exceeded 5\% where all other methods remained well below this limit. 

\begin{table}[htb]   
\begin{center}
	\begin{tabular}{| l | l | l | l | l | l | l | l |}
	\hline
	Expression Quantile & CuffDiff & DESeq & edgeR & limmaQN & limmaVoom & PoissonSeq & baySeq\\ \hline
	100\% (high expression) &  28 & 5 & 3 &  0 & 0 & 7 & 1\\ \hline
   	75\% & 76 & 6 & 0 & 0 & 0& 0 & 0\\ \hline
    	50\% & 84 &27 &1 & 2 & 0& 0 & 0\\ \hline
	25\% (low expression) & 5 & 9 & 0 & 87 & 0& 0 & 0\\ \hline
	Total & 193 & 47 & 4 & 89 & 0 & 7& 1\\
	\hline
	\end{tabular}
\end{center}
	\caption{\footnotesize{\textbf{Number of false DE genes predicted by each method at adjusted p-values (or FDR) $\leq 0.05$ separated by expression quantiles.} Null model p-values were collected from three intra-condition comparisons between replicated libraries of the same origin (see Methods). In total, 16287, 16286, 1620 and 12139 p-values were calculated for genes in the 100\%, 75\%, 50\% and 25\% expression quartiles, respectively. Hence, every gene has three reported p-values from every method representing the three null model comparison. Note that at the bottom 25\% quantile genes with zero counts were excluded.}}
\end{table}

\subsection*{Evaluation of genes expressed in only one condition}
Almost all RNA-seq experiments include a subset of genes that have no detectable read coverage in one of the tested conditions due to silencing. In those cases the assessment of differential expression is confounded by the lack of expression signal in one of the tested conditions that can lead to reduced sensitivity (Type II error), or more commonly to p-values that are inconsistent with the expression levels. Ideally, for this subset of genes the p-values for differential expression should be monotonically correlated with the signal-to-noise ratios ($\mu/\sigma$, ratio of mean over standard deviation) such that higher ratios will be assigned more significant p-values to reflect the confidence in the expression measurement. 

\begin{figure}[!h]
\begin{center}

{\includegraphics[scale=.68]{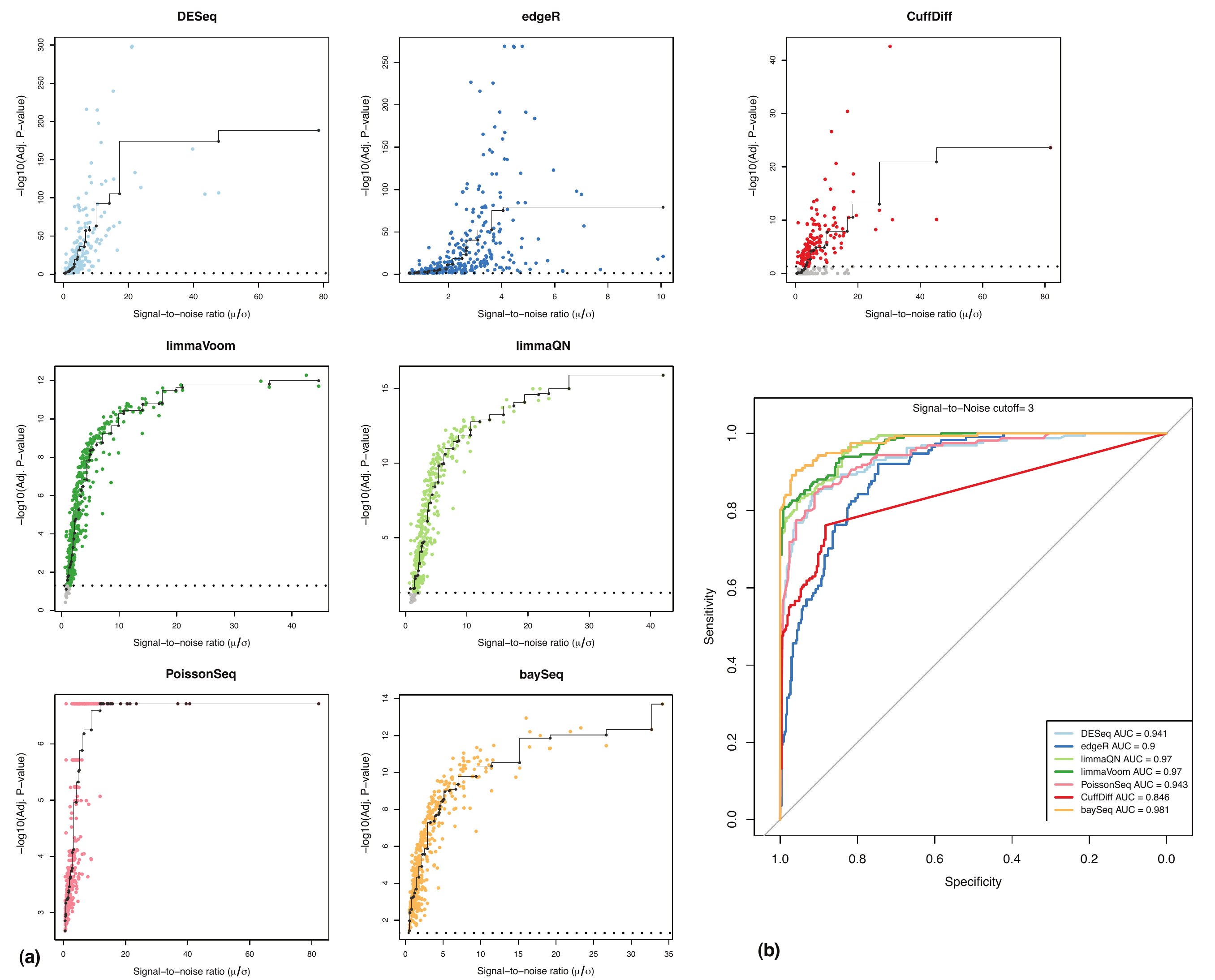}  }
\caption{\footnotesize{(a) Evaluation of the correlation between signal-to-noise and $-log_{10}$(p-value) for 453 genes that were exclusively expressed in condition A or B. Gray shaded points indicate genes with adjusted p-value \textit{larger} than 0.05, which are typically considered not differentially expressed. Cuffdiff, edgeR and DESeq do not properly account for variance in measurements as indicated by poor agreement with the isotonic regression line. (b) ROC curves for detection of DE at signal-to-noise ratio of $\geq3$.}}  \label{ZeroCounts}

\end{center} 
\end{figure}

We evaluated this correlation by an isotonic regression that models the relationship between predictor (signal-to-noise) and response (adjusted p-value) variables with the added constrain of maintaining a monotonic dependency (i.e. if $x_i \leq x_j $ then $f(x_i) \leq f(x_j)$. The results clearly show that the limma and baySeq approaches, and to a large extent PoissonSeq, exhibit the desired monotonic behavior between the signal-to-noise and confidence in differential expression as measured by adjusted p-values (Figure \ref{ZeroCounts}). This was also confirmed by the RMSD between the predicted values from the isotonic regression and the p-values generated by each method. These results indicated a poor correlation for DESeq and edgeR methods (Supplementary Figure 6a). We postulated that for this subset of genes DESeq and edgeR methods default to a Poisson model which implies that the variance is equal to the mean. Hence, the p-values are well correlated with the mean expression (data not shown) but there is no correction for wide variations in gene counts among replicate libraries.

Consistent with the regression analysis the Kendall-tau rank correlation coefficients indicate that limma and baySeq adjusted p-values are best correlated with signal-to-noise whereas Cuffdiff is least correlated although the differences between the methods are less pronounced as indicated by the isotonic regression (Supplementary Figure 6b).
Overall, limma, baySeq, and PoissonSeq had the closest correlation between the two variables demonstrating close to ideal modeling.  

Incorrect modeling of differential expression in this subset of genes may also result in high levels of false negative or false positive predictions where genes with high signal-to-noise ratios are not identified as differentially expressed, or conversely, genes with low signal-to-noise are declared to be differentially expressed. Indeed, with the exception of limma and Cuffdiff, all methods assign adjusted p-values of $\leq 0.05$ to all genes in this data set regardless of their signal-to-noise values. 
To measure the sensitivity and specificity we performed a ROC analysis using a signal-to-noise ratio of $\geq3$ as the classification threshold for differential expression (Figure \ref{ZeroCounts}b). The AUC values support the regression results that limma and baySeq had a performance advantage over other methods. However, both DESeq and PoissonSeq had comparable performance that are close to limma results. Consistent with the rank correlation Cuffdiff showed significantly reduced specificity relative to other methods. This is also illustrated by the large number of false negative genes that have significant signal-to-noise ratios but poor p-values as indicated by the gray points below the 1.3 (i.e. adjusted p-values $>$ 0.05) in figure \ref{ZeroCounts}a. 

\subsection*{Impact of sequencing depth and number of replicate samples on DE detection}
A common challenge when designing RNA-seq experiment is to maximize the detection power of the study under limited budget or sample availability. This has raised a number of practical questions: first, what is the desired sequence coverage for reliable detection of differential expression, and more broadly what is a detection power at a given coverage and number of replicates. Second, given a limited sequencing budget is it preferable to maximize the sequence coverage or increase the number of replicate samples. Finally, what is the impact of different sequencing depths and varying number of replicates on the performances of the DE methods. To address these questions we performed a series of comparisons using combinations of subsets of the sequenced reads and samples. 
We generated a series of down-sampled libraries where a subset of 50\%, 40\%, 30\%, 20\%, 10\% and 5\% reads were randomly sampled from each library (see \nameref{Methods}). This represents a reliable set of varying sequence coverage since any sequencing bias due to nucleotide composition, transcript length, or technical artifact is equally represented in the random sampling. 
We defined the ``true set" of DE genes as the intersection of the DE genes identified by DESeq, edgeR, limmaVoom and baySeq using the full-size libraries and all 5 replicates. We then evaluated DESeq, edgeR, limma and PoissonSeq performance using decreasing number of replicates and sequence coverage for their: i) sensitivity rates, measured as the fraction of the true set, and ii) false positive (FP) rates, defined as the number of genes identified \underline{only} by the evaluated algorithm. 

\begin{figure}[!h]
	\floatpagestyle{empty} 
	\begin{subfigure}[t]{1.0\textwidth} 	
	{\includegraphics[scale=.40]{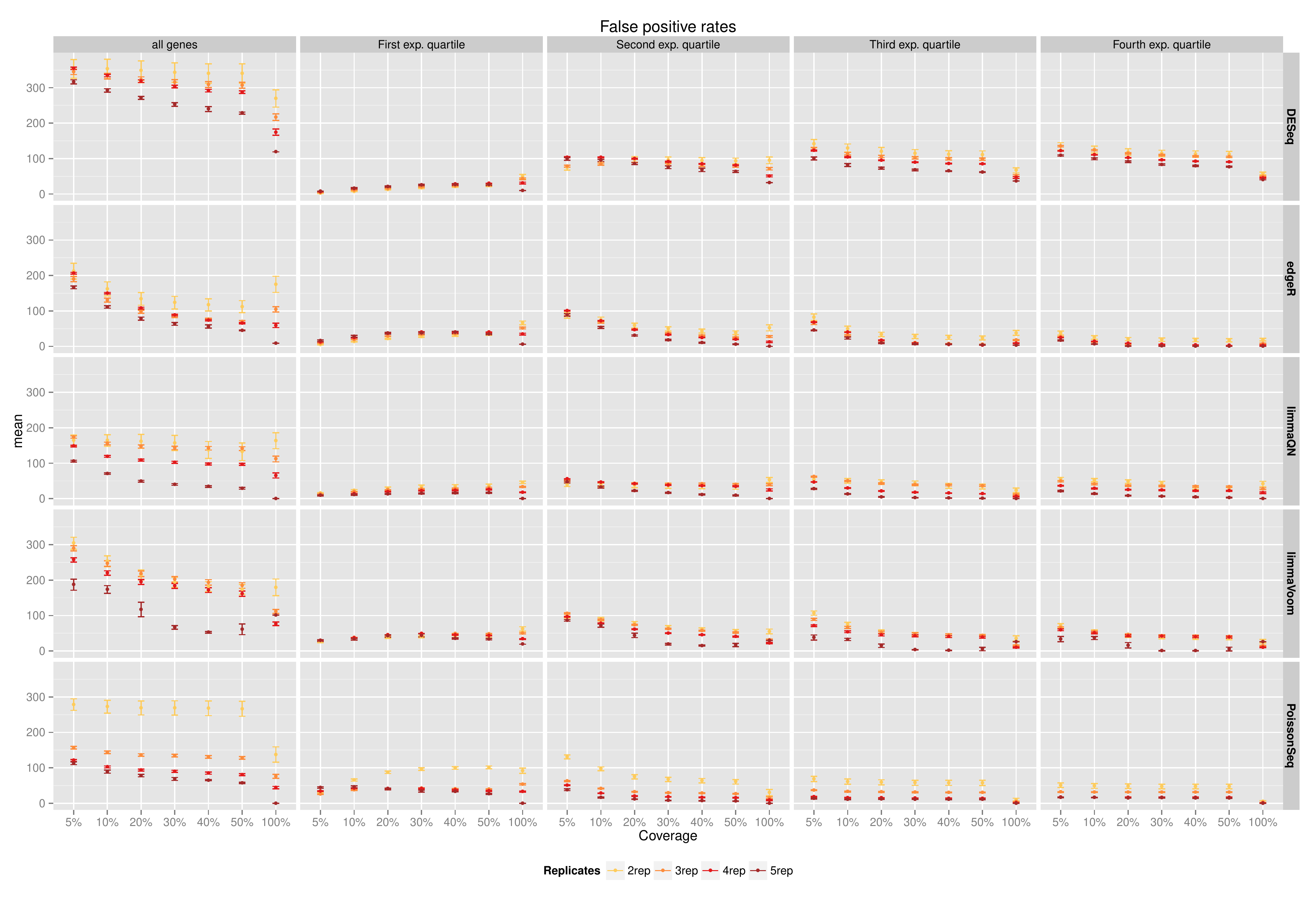}  } 
	\caption{\footnotesize{False positive rates of DE with increasing coverage and sequencing depth}}
	\label{CD:FP}
	\end{subfigure}
	\newpage
	\begin{subfigure}[b]{1.0\textwidth}
	{\includegraphics[scale=.40]{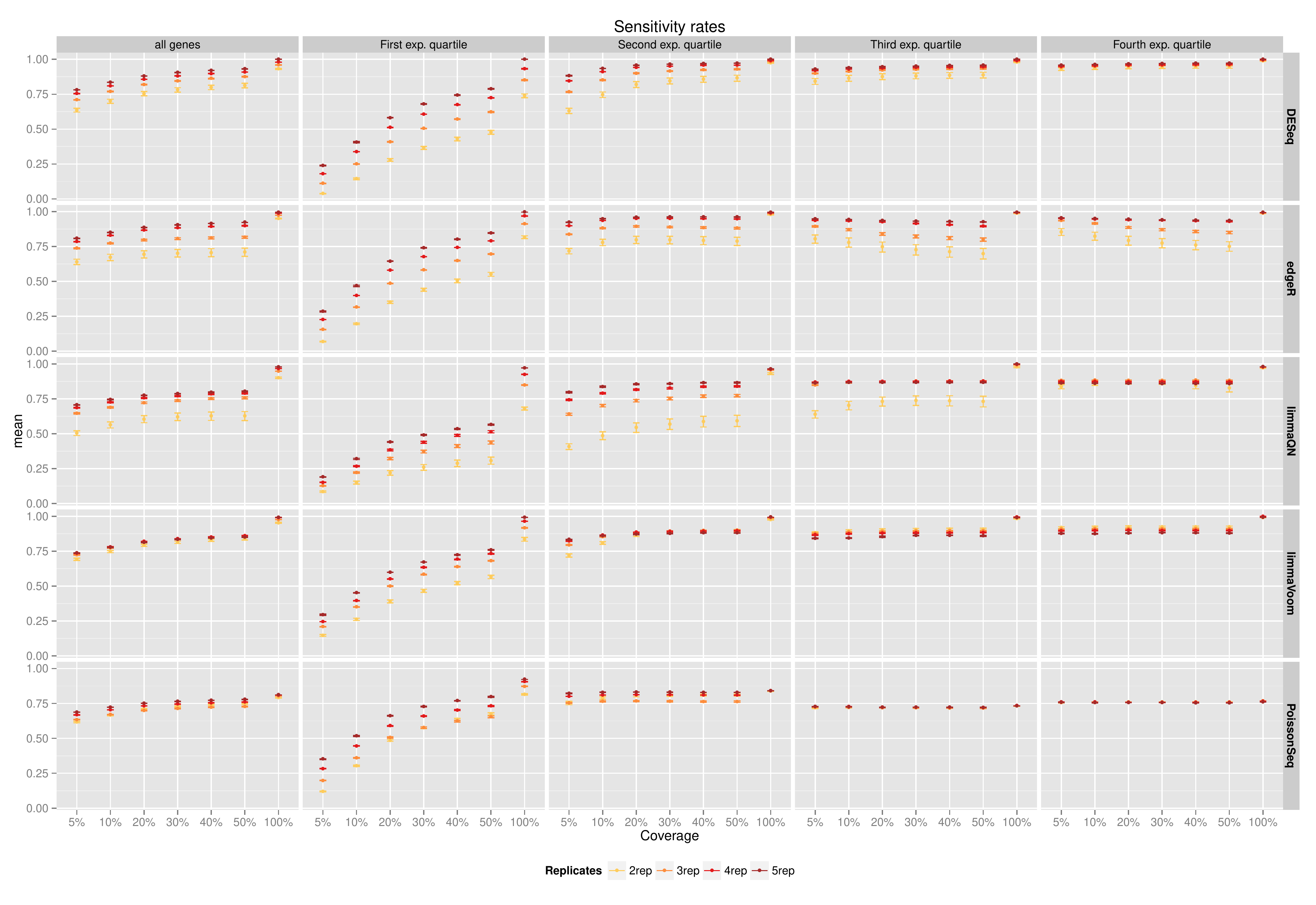}  } 
	\caption{\footnotesize{Sensitivity of DE with increasing coverage and sequencing depth}}
	\label{CD:Sensitivity}
	\end{subfigure}

\caption{\scriptsize{\textbf{Analysis of false positive rates and sensitivity of DE with changing coverage and number of replicate samples.} Analysis performed on all genes as well as on the four expression quartiles. Note that PoissonSeq maximum sensitivity is below 1 since it was not included in the definition of the ``true set".}}  \label{CR}
\end{figure}

As expected, all methods had smaller number of FP with increasing number of replications and increased sequencing depths although there are noticeable differences between the methods. LimmaQN and edgeR had the lowest rates of FP whereas DESeq had the highest (Figure \ref{CR}(a) and Supplementary Figures 7-11). PoissonSeq had the most dramatic reduction in FP rates when increasing the number of replicates from 2 to 3. Interestingly, when examining the FP rates by expression quartiles, the FP rates for the lowest 25\% of expressed genes increased with sequencing depth and number of replicates in contrast to the higher expression quartile where FP is reduced when more data is provided. However, the total number of FP is lowest in the bottom 25\% expression indicating that all methods are conservative when predicting DE at low expression ranges. 

Sensitivity rates also improve significantly with increased sequencing depth and number of replicates although, here as well, significant variabilities exist between methods and between expression levels (Figure \ref{CR}b and Supplementary Figures 7-11). For example, limmaVoom's sensitivity rates are almost independent of the number of replicated samples such that changes in sensitivity are mostly attributed to increasing sequencing depth. This is in contrast to limmQN indicating that normalization has a significant impact on sensitivity to coverage and replication. The most striking impact of coverage and number of replicates is apparent in lowly expressed genes where sensitivity ranges from $<\!10\%$, when comparison is performed with 5\% or reads and two replications, to 100\% detection, when the comparison was performed using the full data set. In contrast, for the highly expressed genes there is little gain in sensitivity with increasing sequencing data or measurements. With most methods, over 90\% of differentially expressed genes at the top expression levels are detected with little as 2 replicates and 5\% of the reads. Surprisingly, edgeR's sensitivity is reduced, for the top half of expressed genes, when sequencing depth increases (Supplementary Figure 8b). Conversely, limmaVoom's sensitivity is reduced for the highly expressed genes with increasing number of replicates (Supplementary Figure 10b).    

Taken together these results lead to two conclusions. First, the number of replicated samples is the most predominant factor in determining DE expression. Second, DE detection of lowly expressed genes is most sensitive to the amount of coverage and replication whereas there is little benefit to increasing sequencing depths for detecting DE in highly expressed genes. 

\section*{Discussion}
In this study we performed a detailed comparative analysis of a number of methods for differential expression analysis from RNA-seq data. Our comparison focused on the performance of the various methods in the areas of normalization, control of false positives, effect of sequencing depth and replication, and on the subset of gene expressed exclusively in one condition. In contrast to other approaches, which rely on simulated data generated by specific statistical distribution \cite{Robles:2012uq,Dillies:2012fk,Kvam:2012fk}, we used the SEQC experimental dataset that had a large fraction of the differentially expressed genes validated by QRT-PCR and includes a cohort of spiked-in controls. 

Overall, no single method emerged as favorable in all comparisons but it is apparent that methods based on the negative binomial modeling have improved specificity and sensitivities as well as good control of false positive errors with comparable performance among DESeq, edgeR, and baySeq.  However, the negative binomial models is not a clear winner in that methods based on other distributions such as PoissonSeq and limma compared favorable to those. On the other hand, Cuffdiff performance has reduced sensitivity and specificity as measured by ROC analysis as well as significant number of false positives in the null model test. We postulate that the source of this is related to the normalization procedure that attempts to account for both alternative isoforms expression and length of transcripts.

Surprisingly, the limma package which was developed and optimized for expression array analysis had comparable, and by some measures improved, performance for both normalization versions tested, relative to the other models that were tailored for RNA-seq analysis. Furthermore, the difference between quantile normalization or the RNA-seq specific \texttt{voom} function in limma was evident in the the number of false positives DE genes in the null model and in the sensitivity to the sequencing depth and number of replicated samples. Limma models the data as a normal distribution which is a reasonable assumption for array intensities but perhaps counterintuitive for count data where it models discrete data with a continuous distribution. However, it is possible that in the limit of large counts it is more important to model the variance accurately than the discreteness. This study demonstrates that for datasets with large number of genes (or tags) the limma package is well suited for detecting DE genes. This suggests that modeling gene count data as log normal distribution, with the appropriate pseudo counts, is a reasonable approximation.   

The results from sequencing depth and replication analysis demonstrate conclusively that the number of sample replicates is the most significant factor in accurate identification of DE genes \cite{Robles:2012uq}. This is not surprising considering that the focus of most methods is to model the variability in gene expression measurements and therefore increased number of replicates adds power to this estimate. Since variability in expression counts decreases with increased mean expression, DE among the highly expressed genes is easily detected even with low sequencing depth and few sample replicates. From a practical point of view studies focus on detecting DE among lowly expressed genes will benefit significantly from increased number of replicates.

Many additional factors that directly impact the detection of differential expression were not considered in this study such as, choice of alignment algorithms, derivation of gene counts, multi-factored studies, detection of alternative transcripts and choice of sequencing platform. Cuffdiff method, for example, incorporates differential isoform detection that is not supported by the simple gene counting methods used here. It is also important to note that the evaluated methods may not be applicable to all RNA-seq data types. For example, small RNA sequencing is not always amenable to quantile normalization as performed in this study (data not shown). Similarly, RNA-seq data from cross-linking and immunoprecipitation (CLIP), or RIP-seq from RNA-binding proteins, are fundamentally different in nature than typical transcriptome profiling and therefore require specialized models. Finally, the field of high-throughput sequencing is rapidly evolving with new technologies continuously introduced. These add additional elements of variability to the measurements that will require specific consideration \cite{Saletore:2012fk}.

The emergence of RNA-seq as the method of choice for transcriptional profiling has motivated the development of a growing number of algorithms for normalization and analysis of this data. This comparative study is the first exhaustive comparison of the widely used DE methods on biologically relevant data. It provides important guidelines for the analysis of RNA-seq data and points the direction for future improvements of RNA-seq analysis.       

\newpage
\bibliographystyle{./text/styles/bmc_article/bmc_article}
\bibliography{./text/RNA-seq_references}


\begin{thebibliography}{10}
\providecommand{\url}[1]{[#1]}
\providecommand{\urlprefix}{}

\bibitem{Mortazavi:2008zr}
Mortazavi A, Williams BA, McCue K, Schaeffer L, Wold B: \textbf{Mapping and
  quantifying mammalian transcriptomes by RNA-Seq}. \emph{Nat Methods} 2008,
  \textbf{5}(7):621--8.

\bibitem{Berger:2010fk}
Berger MF, Levin JZ, Vijayendran K, Sivachenko A, Adiconis X, Maguire J,
  Johnson LA, Robinson J, Verhaak RG, Sougnez C, Onofrio RC, Ziaugra L,
  Cibulskis K, Laine E, Barretina J, Winckler W, Fisher DE, Getz G, Meyerson M,
  Jaffe DB, Gabriel SB, Lander ES, Dummer R, Gnirke A, Nusbaum C, Garraway LA:
  \textbf{Integrative analysis of the melanoma transcriptome}. \emph{Genome
  Res} 2010, \textbf{20}(4):413--27.

\bibitem{Wang:2009kx}
Wang Z, Gerstein M, Snyder M: \textbf{RNA-Seq: a revolutionary tool for
  transcriptomics}. \emph{Nat Rev Genet} 2009, \textbf{10}:57--63.

\bibitem{Young:2012}
Young MD, McCarthy DJ, Wakefield MJ, Smyth GK, Oshlack A, Robinson MD:
  \textbf{Differential Expression for RNA Sequencing (RNA-Seq) Data: Mapping,
  Summarization, Statistical Analysis, and Experimental Design}. In
  \emph{Bioinformatics for High Throughput Sequencing}. Edited by
  Rodr{\'\i}guez-Ezpeleta N, Hackenberg M, Aransay AM, Springer New York
  2012:169--190,
  \urlprefix\url{[http://dx.doi.org/10.1007/978-1-4614-0782-9_10]}.

\bibitem{Trapnell:2010ve}
Trapnell C, Williams BA, Pertea G, Mortazavi A, Kwan G, van Baren MJ, Salzberg
  SL, Wold BJ, Pachter L: \textbf{Transcript assembly and quantification by
  RNA-Seq reveals unannotated transcripts and isoform switching during cell
  differentiation}. \emph{Nat Biotechnol} 2010, \textbf{28}(5):511--5.

\bibitem{Robinson:2010fk}
Robinson MD, McCarthy DJ, Smyth GK: \textbf{edgeR: a Bioconductor package for
  differential expression analysis of digital gene expression data}.
  \emph{Bioinformatics} 2010, \textbf{26}:139--40.

\bibitem{Anders:2010uq}
Anders S, Huber W: \textbf{Differential expression analysis for sequence count
  data}. \emph{Genome Biol} 2010, \textbf{11}(10):R106.

\bibitem{Li:2012ly}
Li J, Witten DM, Johnstone IM, Tibshirani R: \textbf{Normalization, testing,
  and false discovery rate estimation for RNA-sequencing data}.
  \emph{Biostatistics} 2012, \textbf{13}(3):523--38.

\bibitem{Hardcastle:2010fk}
Hardcastle TJ, Kelly KA: \textbf{baySeq: empirical Bayesian methods for
  identifying differential expression in sequence count data}. \emph{BMC
  Bioinformatics} 2010, \textbf{11}:422.

\bibitem{Smyth:2004nx}
Smyth GK: \textbf{Linear models and empirical bayes methods for assessing
  differential expression in microarray experiments}. \emph{Stat Appl Genet Mol
  Biol} 2004, \textbf{3}:Article3.

\bibitem{Shi:2010vn}
Shi L, Campbell G, Jones WD, Campagne F, Wen Z, Walker SJ, Su Z, Chu TM,
  Goodsaid FM, Pusztai L, Shaughnessy JD Jr, Oberthuer A, Thomas RS, Paules RS,
  Fielden M, Barlogie B, Chen W, Du P, Fischer M, Furlanello C, Gallas BD, Ge
  X, Megherbi DB, Symmans WF, Wang MD, Zhang J, Bitter H, Brors B, Bushel PR,
  Bylesjo M, Chen M, Cheng J, Cheng J, Chou J, Davison TS, Delorenzi M, Deng Y,
  Devanarayan V, Dix DJ, Dopazo J, Dorff KC, Elloumi F, Fan J, Fan S, Fan X,
  Fang H, Gonzaludo N, Hess KR, Hong H, Huan J, Irizarry RA, Judson R, Juraeva
  D, Lababidi S, Lambert CG, Li L, Li Y, Li Z, Lin SM, Liu G, Lobenhofer EK,
  Luo J, Luo W, McCall MN, Nikolsky Y, Pennello GA, Perkins RG, Philip R,
  Popovici V, Price ND, Qian F, Scherer A, Shi T, Shi W, Sung J, Thierry-Mieg
  D, Thierry-Mieg J, Thodima V, Trygg J, Vishnuvajjala L, Wang SJ, Wu J, Wu Y,
  Xie Q, Yousef WA, Zhang L, Zhang X, Zhong S, Zhou Y, Zhu S, Arasappan D, Bao
  W, Lucas AB, Berthold F, Brennan RJ, Buness A, Catalano JG, Chang C, Chen R,
  Cheng Y, Cui J, Czika W, Demichelis F, Deng X, Dosymbekov D, Eils R, Feng Y,
  Fostel J, Fulmer-Smentek S, Fuscoe JC, Gatto L, Ge W, Goldstein DR, Guo L,
  Halbert DN, Han J, Harris SC, Hatzis C, Herman D, Huang J, Jensen RV, Jiang
  R, Johnson CD, Jurman G, Kahlert Y, Khuder SA, Kohl M, Li J, Li L, Li M, Li
  QZ, Li S, Li Z, Liu J, Liu Y, Liu Z, Meng L, Madera M, Martinez-Murillo F,
  Medina I, Meehan J, Miclaus K, Moffitt RA, Montaner D, Mukherjee P, Mulligan
  GJ, Neville P, Nikolskaya T, Ning B, Page GP, Parker J, Parry RM, Peng X,
  Peterson RL, Phan JH, Quanz B, Ren Y, Riccadonna S, Roter AH, Samuelson FW,
  Schumacher MM, Shambaugh JD, Shi Q, Shippy R, Si S, Smalter A, Sotiriou C,
  Soukup M, Staedtler F, Steiner G, Stokes TH, Sun Q, Tan PY, Tang R, Tezak Z,
  Thorn B, Tsyganova M, Turpaz Y, Vega SC, Visintainer R, von Frese J, Wang C,
  Wang E, Wang J, Wang W, Westermann F, Willey JC, Woods M, Wu S, Xiao N, Xu J,
  Xu L, Yang L, Zeng X, Zhang J, Zhang L, Zhang M, Zhao C, Puri RK, Scherf U,
  Tong W, Wolfinger RD, {MAQC Consortium}: \textbf{The MicroArray Quality
  Control (MAQC)-II study of common practices for the development and
  validation of microarray-based predictive models}. \emph{Nat Biotechnol}
  2010, \textbf{28}(8):827--38.

\bibitem{HTseq}
Anders S: \textbf{HTSeq: Analysis of high-throughput sequencing data with
  Python.}  2011,
  \urlprefix\url{[http://www-huber.embl.de/users/anders/HTSeq/]}.

\bibitem{picard}
Wysoker A, Tibbetts K, Fennell T: \textbf{Picard}  2012,
  \urlprefix\url{[http://picard.sourceforge.net/]}.

\bibitem{Quinlan:2010fk}
Quinlan AR, Hall IM: \textbf{BEDTools: a flexible suite of utilities for
  comparing genomic features}. \emph{Bioinformatics} 2010,
  \textbf{26}(6):841--2.

\bibitem{Dillies:2012fk}
Dillies MA, Rau A, Aubert J, Hennequet-Antier C, Jeanmougin M, Servant N, Keime
  C, Marot G, Castel D, Estelle J, Guernec G, Jagla B, Jouneau L, Lalo{\"e} D,
  Le~Gall C, Scha{\"e}ffer B, Le~Crom S, Guedj M, Jaffr{\'e}zic F, {on behalf
  of The French StatOmique Consortium}: \textbf{A comprehensive evaluation of
  normalization methods for Illumina high-throughput RNA sequencing data
  analysis}. \emph{Brief Bioinform} 2012.

\bibitem{Bullard:2010ys}
Bullard JH, Purdom E, Hansen KD, Dudoit S: \textbf{Evaluation of statistical
  methods for normalization and differential expression in mRNA-Seq
  experiments}. \emph{BMC Bioinformatics} 2010, \textbf{11}:94.

\bibitem{Robinson:2010kl}
Robinson MD, Oshlack A: \textbf{A scaling normalization method for differential
  expression analysis of RNA-seq data}. \emph{Genome Biol} 2010,
  \textbf{11}(3):R25.

\bibitem{Bolstad:2003fk}
Bolstad BM, Irizarry RA, Astrand M, Speed TP: \textbf{A comparison of
  normalization methods for high density oligonucleotide array data based on
  variance and bias}. \emph{Bioinformatics} 2003, \textbf{19}(2):185--93.

\bibitem{Robinson:2007bh}
Robinson MD, Smyth GK: \textbf{Moderated statistical tests for assessing
  differences in tag abundance}. \emph{Bioinformatics} 2007,
  \textbf{23}(21):2881--7.

\bibitem{Nagalakshmi:2008dq}
Nagalakshmi U, Wang Z, Waern K, Shou C, Raha D, Gerstein M, Snyder M:
  \textbf{The transcriptional landscape of the yeast genome defined by RNA
  sequencing}. \emph{Science} 2008, \textbf{320}(5881):1344--9.

\bibitem{Marioni:2008oq}
Marioni JC, Mason CE, Mane SM, Stephens M, Gilad Y: \textbf{RNA-seq: an
  assessment of technical reproducibility and comparison with gene expression
  arrays}. \emph{Genome Res} 2008, \textbf{18}(9):1509--17.

\bibitem{Trapnell:2009uq}
Trapnell C, Pachter L, Salzberg SL: \textbf{TopHat: discovering splice
  junctions with RNA-Seq}. \emph{Bioinformatics} 2009, \textbf{25}(9):1105--11.

\bibitem{Canales:2006fk}
Canales RD, Luo Y, Willey JC, Austermiller B, Barbacioru CC, Boysen C,
  Hunkapiller K, Jensen RV, Knight CR, Lee KY, Ma Y, Maqsodi B, Papallo A,
  Peters EH, Poulter K, Ruppel PL, Samaha RR, Shi L, Yang W, Zhang L, Goodsaid
  FM: \textbf{Evaluation of DNA microarray results with quantitative gene
  expression platforms}. \emph{Nat Biotechnol} 2006, \textbf{24}(9):1115--22.

\bibitem{Anders:2012fk}
Anders S, Reyes A, Huber W: \textbf{Detecting differential usage of exons from
  RNA-seq data}. \emph{Genome Res} 2012, \textbf{22}(10):2008--17.

\bibitem{Robles:2012uq}
Robles JA, Qureshi SE, Stephen SJ, Wilson SR, Burden CJ, Taylor JM:
  \textbf{Efficient experimental design and analysis strategies for the
  detection of differential expression using RNA-Sequencing}. \emph{BMC
  Genomics} 2012, \textbf{13}:484.

\bibitem{Kvam:2012fk}
Kvam VM, Liu P, Si Y: \textbf{A comparison of statistical methods for detecting
  differentially expressed genes from RNA-seq data}. \emph{Am J Bot} 2012,
  \textbf{99}(2):248--56.

\bibitem{Saletore:2012fk}
Saletore Y, Meyer K, Korlach J, Vilfan ID, Jaffrey S, Mason CE: \textbf{The
  birth of the Epitranscriptome: deciphering the function of RNA
  modifications}. \emph{Genome Biol} 2012, \textbf{13}(10):175.

\end{thebibliography}

\newcommand{\BMCxmlcomment}[1]{}

\BMCxmlcomment{

<refgrp>

<bibl id="B1">
  <title><p>Mapping and quantifying mammalian transcriptomes by
  RNA-Seq</p></title>
  <aug>
    <au><snm>Mortazavi</snm><fnm>A</fnm></au>
    <au><snm>Williams</snm><fnm>BA</fnm></au>
    <au><snm>McCue</snm><fnm>K</fnm></au>
    <au><snm>Schaeffer</snm><fnm>L</fnm></au>
    <au><snm>Wold</snm><fnm>B</fnm></au>
  </aug>
  <source>Nat Methods</source>
  <pubdate>2008</pubdate>
  <volume>5</volume>
  <issue>7</issue>
  <fpage>621</fpage>
  <lpage>8</lpage>
</bibl>

<bibl id="B2">
  <title><p>Integrative analysis of the melanoma transcriptome</p></title>
  <aug>
    <au><snm>Berger</snm><fnm>MF</fnm></au>
    <au><snm>Levin</snm><fnm>JZ</fnm></au>
    <au><snm>Vijayendran</snm><fnm>K</fnm></au>
    <au><snm>Sivachenko</snm><fnm>A</fnm></au>
    <au><snm>Adiconis</snm><fnm>X</fnm></au>
    <au><snm>Maguire</snm><fnm>J</fnm></au>
    <au><snm>Johnson</snm><fnm>LA</fnm></au>
    <au><snm>Robinson</snm><fnm>J</fnm></au>
    <au><snm>Verhaak</snm><fnm>RG</fnm></au>
    <au><snm>Sougnez</snm><fnm>C</fnm></au>
    <au><snm>Onofrio</snm><fnm>RC</fnm></au>
    <au><snm>Ziaugra</snm><fnm>L</fnm></au>
    <au><snm>Cibulskis</snm><fnm>K</fnm></au>
    <au><snm>Laine</snm><fnm>E</fnm></au>
    <au><snm>Barretina</snm><fnm>J</fnm></au>
    <au><snm>Winckler</snm><fnm>W</fnm></au>
    <au><snm>Fisher</snm><fnm>DE</fnm></au>
    <au><snm>Getz</snm><fnm>G</fnm></au>
    <au><snm>Meyerson</snm><fnm>M</fnm></au>
    <au><snm>Jaffe</snm><fnm>DB</fnm></au>
    <au><snm>Gabriel</snm><fnm>SB</fnm></au>
    <au><snm>Lander</snm><fnm>ES</fnm></au>
    <au><snm>Dummer</snm><fnm>R</fnm></au>
    <au><snm>Gnirke</snm><fnm>A</fnm></au>
    <au><snm>Nusbaum</snm><fnm>C</fnm></au>
    <au><snm>Garraway</snm><fnm>LA</fnm></au>
  </aug>
  <source>Genome Res</source>
  <pubdate>2010</pubdate>
  <volume>20</volume>
  <issue>4</issue>
  <fpage>413</fpage>
  <lpage>27</lpage>
</bibl>

<bibl id="B3">
  <title><p>RNA-Seq: a revolutionary tool for transcriptomics</p></title>
  <aug>
    <au><snm>Wang</snm><fnm>Z</fnm></au>
    <au><snm>Gerstein</snm><fnm>M</fnm></au>
    <au><snm>Snyder</snm><fnm>M</fnm></au>
  </aug>
  <source>Nat Rev Genet</source>
  <pubdate>2009</pubdate>
  <volume>10</volume>
  <issue>1</issue>
  <fpage>57</fpage>
  <lpage>63</lpage>
</bibl>

<bibl id="B4">
  <title><p>Differential Expression for RNA Sequencing (RNA-Seq) Data: Mapping,
  Summarization, Statistical Analysis, and Experimental Design</p></title>
  <aug>
    <au><snm>Young</snm><fnm>MD</fnm></au>
    <au><snm>McCarthy</snm><fnm>DJ</fnm></au>
    <au><snm>Wakefield</snm><fnm>MJ</fnm></au>
    <au><snm>Smyth</snm><fnm>GK</fnm></au>
    <au><snm>Oshlack</snm><fnm>A</fnm></au>
    <au><snm>Robinson</snm><fnm>MD</fnm></au>
  </aug>
  <source>Bioinformatics for High Throughput Sequencing</source>
  <publisher>Springer New York</publisher>
  <editor>Rodr{\'\i}guez-Ezpeleta, Naiara and Hackenberg, Michael and Aransay,
  Ana M.</editor>
  <pubdate>2012</pubdate>
  <fpage>169</fpage>
  <lpage>190</lpage>
  <url>http://dx.doi.org/10.1007/978-1-4614-0782-9_10</url>
</bibl>

<bibl id="B5">
  <title><p>Transcript assembly and quantification by RNA-Seq reveals
  unannotated transcripts and isoform switching during cell
  differentiation</p></title>
  <aug>
    <au><snm>Trapnell</snm><fnm>C</fnm></au>
    <au><snm>Williams</snm><fnm>BA</fnm></au>
    <au><snm>Pertea</snm><fnm>G</fnm></au>
    <au><snm>Mortazavi</snm><fnm>A</fnm></au>
    <au><snm>Kwan</snm><fnm>G</fnm></au>
    <au><snm>Baren</snm><fnm>MJ</fnm></au>
    <au><snm>Salzberg</snm><fnm>SL</fnm></au>
    <au><snm>Wold</snm><fnm>BJ</fnm></au>
    <au><snm>Pachter</snm><fnm>L</fnm></au>
  </aug>
  <source>Nat Biotechnol</source>
  <pubdate>2010</pubdate>
  <volume>28</volume>
  <issue>5</issue>
  <fpage>511</fpage>
  <lpage>5</lpage>
</bibl>

<bibl id="B6">
  <title><p>edgeR: a Bioconductor package for differential expression analysis
  of digital gene expression data</p></title>
  <aug>
    <au><snm>Robinson</snm><fnm>MD</fnm></au>
    <au><snm>McCarthy</snm><fnm>DJ</fnm></au>
    <au><snm>Smyth</snm><fnm>GK</fnm></au>
  </aug>
  <source>Bioinformatics</source>
  <pubdate>2010</pubdate>
  <volume>26</volume>
  <issue>1</issue>
  <fpage>139</fpage>
  <lpage>40</lpage>
</bibl>

<bibl id="B7">
  <title><p>Differential expression analysis for sequence count
  data</p></title>
  <aug>
    <au><snm>Anders</snm><fnm>S</fnm></au>
    <au><snm>Huber</snm><fnm>W</fnm></au>
  </aug>
  <source>Genome Biol</source>
  <pubdate>2010</pubdate>
  <volume>11</volume>
  <issue>10</issue>
  <fpage>R106</fpage>
</bibl>

<bibl id="B8">
  <title><p>Normalization, testing, and false discovery rate estimation for
  RNA-sequencing data</p></title>
  <aug>
    <au><snm>Li</snm><fnm>J</fnm></au>
    <au><snm>Witten</snm><fnm>DM</fnm></au>
    <au><snm>Johnstone</snm><fnm>IM</fnm></au>
    <au><snm>Tibshirani</snm><fnm>R</fnm></au>
  </aug>
  <source>Biostatistics</source>
  <pubdate>2012</pubdate>
  <volume>13</volume>
  <issue>3</issue>
  <fpage>523</fpage>
  <lpage>38</lpage>
</bibl>

<bibl id="B9">
  <title><p>baySeq: empirical Bayesian methods for identifying differential
  expression in sequence count data</p></title>
  <aug>
    <au><snm>Hardcastle</snm><fnm>TJ</fnm></au>
    <au><snm>Kelly</snm><fnm>KA</fnm></au>
  </aug>
  <source>BMC Bioinformatics</source>
  <pubdate>2010</pubdate>
  <volume>11</volume>
  <fpage>422</fpage>
</bibl>

<bibl id="B10">
  <title><p>Linear models and empirical bayes methods for assessing
  differential expression in microarray experiments</p></title>
  <aug>
    <au><snm>Smyth</snm><fnm>GK</fnm></au>
  </aug>
  <source>Stat Appl Genet Mol Biol</source>
  <pubdate>2004</pubdate>
  <volume>3</volume>
  <fpage>Article3</fpage>
</bibl>

<bibl id="B11">
  <title><p>The MicroArray Quality Control (MAQC)-II study of common practices
  for the development and validation of microarray-based predictive
  models</p></title>
  <aug>
    <au><snm>Shi</snm><fnm>L</fnm></au>
    <au><snm>Campbell</snm><fnm>G</fnm></au>
    <au><snm>Jones</snm><fnm>WD</fnm></au>
    <au><snm>Campagne</snm><fnm>F</fnm></au>
    <au><snm>Wen</snm><fnm>Z</fnm></au>
    <au><snm>Walker</snm><fnm>SJ</fnm></au>
    <au><snm>Su</snm><fnm>Z</fnm></au>
    <au><snm>Chu</snm><fnm>TM</fnm></au>
    <au><snm>Goodsaid</snm><fnm>FM</fnm></au>
    <au><snm>Pusztai</snm><fnm>L</fnm></au>
    <au><snm>Shaughnessy</snm><fnm>JD</fnm></au>
    <au><snm>Oberthuer</snm><fnm>A</fnm></au>
    <au><snm>Thomas</snm><fnm>RS</fnm></au>
    <au><snm>Paules</snm><fnm>RS</fnm></au>
    <au><snm>Fielden</snm><fnm>M</fnm></au>
    <au><snm>Barlogie</snm><fnm>B</fnm></au>
    <au><snm>Chen</snm><fnm>W</fnm></au>
    <au><snm>Du</snm><fnm>P</fnm></au>
    <au><snm>Fischer</snm><fnm>M</fnm></au>
    <au><snm>Furlanello</snm><fnm>C</fnm></au>
    <au><snm>Gallas</snm><fnm>BD</fnm></au>
    <au><snm>Ge</snm><fnm>X</fnm></au>
    <au><snm>Megherbi</snm><fnm>DB</fnm></au>
    <au><snm>Symmans</snm><fnm>WF</fnm></au>
    <au><snm>Wang</snm><fnm>MD</fnm></au>
    <au><snm>Zhang</snm><fnm>J</fnm></au>
    <au><snm>Bitter</snm><fnm>H</fnm></au>
    <au><snm>Brors</snm><fnm>B</fnm></au>
    <au><snm>Bushel</snm><fnm>PR</fnm></au>
    <au><snm>Bylesjo</snm><fnm>M</fnm></au>
    <au><snm>Chen</snm><fnm>M</fnm></au>
    <au><snm>Cheng</snm><fnm>J</fnm></au>
    <au><snm>Cheng</snm><fnm>J</fnm></au>
    <au><snm>Chou</snm><fnm>J</fnm></au>
    <au><snm>Davison</snm><fnm>TS</fnm></au>
    <au><snm>Delorenzi</snm><fnm>M</fnm></au>
    <au><snm>Deng</snm><fnm>Y</fnm></au>
    <au><snm>Devanarayan</snm><fnm>V</fnm></au>
    <au><snm>Dix</snm><fnm>DJ</fnm></au>
    <au><snm>Dopazo</snm><fnm>J</fnm></au>
    <au><snm>Dorff</snm><fnm>KC</fnm></au>
    <au><snm>Elloumi</snm><fnm>F</fnm></au>
    <au><snm>Fan</snm><fnm>J</fnm></au>
    <au><snm>Fan</snm><fnm>S</fnm></au>
    <au><snm>Fan</snm><fnm>X</fnm></au>
    <au><snm>Fang</snm><fnm>H</fnm></au>
    <au><snm>Gonzaludo</snm><fnm>N</fnm></au>
    <au><snm>Hess</snm><fnm>KR</fnm></au>
    <au><snm>Hong</snm><fnm>H</fnm></au>
    <au><snm>Huan</snm><fnm>J</fnm></au>
    <au><snm>Irizarry</snm><fnm>RA</fnm></au>
    <au><snm>Judson</snm><fnm>R</fnm></au>
    <au><snm>Juraeva</snm><fnm>D</fnm></au>
    <au><snm>Lababidi</snm><fnm>S</fnm></au>
    <au><snm>Lambert</snm><fnm>CG</fnm></au>
    <au><snm>Li</snm><fnm>L</fnm></au>
    <au><snm>Li</snm><fnm>Y</fnm></au>
    <au><snm>Li</snm><fnm>Z</fnm></au>
    <au><snm>Lin</snm><fnm>SM</fnm></au>
    <au><snm>Liu</snm><fnm>G</fnm></au>
    <au><snm>Lobenhofer</snm><fnm>EK</fnm></au>
    <au><snm>Luo</snm><fnm>J</fnm></au>
    <au><snm>Luo</snm><fnm>W</fnm></au>
    <au><snm>McCall</snm><fnm>MN</fnm></au>
    <au><snm>Nikolsky</snm><fnm>Y</fnm></au>
    <au><snm>Pennello</snm><fnm>GA</fnm></au>
    <au><snm>Perkins</snm><fnm>RG</fnm></au>
    <au><snm>Philip</snm><fnm>R</fnm></au>
    <au><snm>Popovici</snm><fnm>V</fnm></au>
    <au><snm>Price</snm><fnm>ND</fnm></au>
    <au><snm>Qian</snm><fnm>F</fnm></au>
    <au><snm>Scherer</snm><fnm>A</fnm></au>
    <au><snm>Shi</snm><fnm>T</fnm></au>
    <au><snm>Shi</snm><fnm>W</fnm></au>
    <au><snm>Sung</snm><fnm>J</fnm></au>
    <au><snm>Thierry Mieg</snm><fnm>D</fnm></au>
    <au><snm>Thierry Mieg</snm><fnm>J</fnm></au>
    <au><snm>Thodima</snm><fnm>V</fnm></au>
    <au><snm>Trygg</snm><fnm>J</fnm></au>
    <au><snm>Vishnuvajjala</snm><fnm>L</fnm></au>
    <au><snm>Wang</snm><fnm>SJ</fnm></au>
    <au><snm>Wu</snm><fnm>J</fnm></au>
    <au><snm>Wu</snm><fnm>Y</fnm></au>
    <au><snm>Xie</snm><fnm>Q</fnm></au>
    <au><snm>Yousef</snm><fnm>WA</fnm></au>
    <au><snm>Zhang</snm><fnm>L</fnm></au>
    <au><snm>Zhang</snm><fnm>X</fnm></au>
    <au><snm>Zhong</snm><fnm>S</fnm></au>
    <au><snm>Zhou</snm><fnm>Y</fnm></au>
    <au><snm>Zhu</snm><fnm>S</fnm></au>
    <au><snm>Arasappan</snm><fnm>D</fnm></au>
    <au><snm>Bao</snm><fnm>W</fnm></au>
    <au><snm>Lucas</snm><fnm>AB</fnm></au>
    <au><snm>Berthold</snm><fnm>F</fnm></au>
    <au><snm>Brennan</snm><fnm>RJ</fnm></au>
    <au><snm>Buness</snm><fnm>A</fnm></au>
    <au><snm>Catalano</snm><fnm>JG</fnm></au>
    <au><snm>Chang</snm><fnm>C</fnm></au>
    <au><snm>Chen</snm><fnm>R</fnm></au>
    <au><snm>Cheng</snm><fnm>Y</fnm></au>
    <au><snm>Cui</snm><fnm>J</fnm></au>
    <au><snm>Czika</snm><fnm>W</fnm></au>
    <au><snm>Demichelis</snm><fnm>F</fnm></au>
    <au><snm>Deng</snm><fnm>X</fnm></au>
    <au><snm>Dosymbekov</snm><fnm>D</fnm></au>
    <au><snm>Eils</snm><fnm>R</fnm></au>
    <au><snm>Feng</snm><fnm>Y</fnm></au>
    <au><snm>Fostel</snm><fnm>J</fnm></au>
    <au><snm>Fulmer Smentek</snm><fnm>S</fnm></au>
    <au><snm>Fuscoe</snm><fnm>JC</fnm></au>
    <au><snm>Gatto</snm><fnm>L</fnm></au>
    <au><snm>Ge</snm><fnm>W</fnm></au>
    <au><snm>Goldstein</snm><fnm>DR</fnm></au>
    <au><snm>Guo</snm><fnm>L</fnm></au>
    <au><snm>Halbert</snm><fnm>DN</fnm></au>
    <au><snm>Han</snm><fnm>J</fnm></au>
    <au><snm>Harris</snm><fnm>SC</fnm></au>
    <au><snm>Hatzis</snm><fnm>C</fnm></au>
    <au><snm>Herman</snm><fnm>D</fnm></au>
    <au><snm>Huang</snm><fnm>J</fnm></au>
    <au><snm>Jensen</snm><fnm>RV</fnm></au>
    <au><snm>Jiang</snm><fnm>R</fnm></au>
    <au><snm>Johnson</snm><fnm>CD</fnm></au>
    <au><snm>Jurman</snm><fnm>G</fnm></au>
    <au><snm>Kahlert</snm><fnm>Y</fnm></au>
    <au><snm>Khuder</snm><fnm>SA</fnm></au>
    <au><snm>Kohl</snm><fnm>M</fnm></au>
    <au><snm>Li</snm><fnm>J</fnm></au>
    <au><snm>Li</snm><fnm>L</fnm></au>
    <au><snm>Li</snm><fnm>M</fnm></au>
    <au><snm>Li</snm><fnm>QZ</fnm></au>
    <au><snm>Li</snm><fnm>S</fnm></au>
    <au><snm>Li</snm><fnm>Z</fnm></au>
    <au><snm>Liu</snm><fnm>J</fnm></au>
    <au><snm>Liu</snm><fnm>Y</fnm></au>
    <au><snm>Liu</snm><fnm>Z</fnm></au>
    <au><snm>Meng</snm><fnm>L</fnm></au>
    <au><snm>Madera</snm><fnm>M</fnm></au>
    <au><snm>Martinez Murillo</snm><fnm>F</fnm></au>
    <au><snm>Medina</snm><fnm>I</fnm></au>
    <au><snm>Meehan</snm><fnm>J</fnm></au>
    <au><snm>Miclaus</snm><fnm>K</fnm></au>
    <au><snm>Moffitt</snm><fnm>RA</fnm></au>
    <au><snm>Montaner</snm><fnm>D</fnm></au>
    <au><snm>Mukherjee</snm><fnm>P</fnm></au>
    <au><snm>Mulligan</snm><fnm>GJ</fnm></au>
    <au><snm>Neville</snm><fnm>P</fnm></au>
    <au><snm>Nikolskaya</snm><fnm>T</fnm></au>
    <au><snm>Ning</snm><fnm>B</fnm></au>
    <au><snm>Page</snm><fnm>GP</fnm></au>
    <au><snm>Parker</snm><fnm>J</fnm></au>
    <au><snm>Parry</snm><fnm>RM</fnm></au>
    <au><snm>Peng</snm><fnm>X</fnm></au>
    <au><snm>Peterson</snm><fnm>RL</fnm></au>
    <au><snm>Phan</snm><fnm>JH</fnm></au>
    <au><snm>Quanz</snm><fnm>B</fnm></au>
    <au><snm>Ren</snm><fnm>Y</fnm></au>
    <au><snm>Riccadonna</snm><fnm>S</fnm></au>
    <au><snm>Roter</snm><fnm>AH</fnm></au>
    <au><snm>Samuelson</snm><fnm>FW</fnm></au>
    <au><snm>Schumacher</snm><fnm>MM</fnm></au>
    <au><snm>Shambaugh</snm><fnm>JD</fnm></au>
    <au><snm>Shi</snm><fnm>Q</fnm></au>
    <au><snm>Shippy</snm><fnm>R</fnm></au>
    <au><snm>Si</snm><fnm>S</fnm></au>
    <au><snm>Smalter</snm><fnm>A</fnm></au>
    <au><snm>Sotiriou</snm><fnm>C</fnm></au>
    <au><snm>Soukup</snm><fnm>M</fnm></au>
    <au><snm>Staedtler</snm><fnm>F</fnm></au>
    <au><snm>Steiner</snm><fnm>G</fnm></au>
    <au><snm>Stokes</snm><fnm>TH</fnm></au>
    <au><snm>Sun</snm><fnm>Q</fnm></au>
    <au><snm>Tan</snm><fnm>PY</fnm></au>
    <au><snm>Tang</snm><fnm>R</fnm></au>
    <au><snm>Tezak</snm><fnm>Z</fnm></au>
    <au><snm>Thorn</snm><fnm>B</fnm></au>
    <au><snm>Tsyganova</snm><fnm>M</fnm></au>
    <au><snm>Turpaz</snm><fnm>Y</fnm></au>
    <au><snm>Vega</snm><fnm>SC</fnm></au>
    <au><snm>Visintainer</snm><fnm>R</fnm></au>
    <au><snm>Frese</snm><fnm>J</fnm></au>
    <au><snm>Wang</snm><fnm>C</fnm></au>
    <au><snm>Wang</snm><fnm>E</fnm></au>
    <au><snm>Wang</snm><fnm>J</fnm></au>
    <au><snm>Wang</snm><fnm>W</fnm></au>
    <au><snm>Westermann</snm><fnm>F</fnm></au>
    <au><snm>Willey</snm><fnm>JC</fnm></au>
    <au><snm>Woods</snm><fnm>M</fnm></au>
    <au><snm>Wu</snm><fnm>S</fnm></au>
    <au><snm>Xiao</snm><fnm>N</fnm></au>
    <au><snm>Xu</snm><fnm>J</fnm></au>
    <au><snm>Xu</snm><fnm>L</fnm></au>
    <au><snm>Yang</snm><fnm>L</fnm></au>
    <au><snm>Zeng</snm><fnm>X</fnm></au>
    <au><snm>Zhang</snm><fnm>J</fnm></au>
    <au><snm>Zhang</snm><fnm>L</fnm></au>
    <au><snm>Zhang</snm><fnm>M</fnm></au>
    <au><snm>Zhao</snm><fnm>C</fnm></au>
    <au><snm>Puri</snm><fnm>RK</fnm></au>
    <au><snm>Scherf</snm><fnm>U</fnm></au>
    <au><snm>Tong</snm><fnm>W</fnm></au>
    <au><snm>Wolfinger</snm><fnm>RD</fnm></au>
    <au><cnm>{MAQC Consortium}</cnm></au>
  </aug>
  <source>Nat Biotechnol</source>
  <pubdate>2010</pubdate>
  <volume>28</volume>
  <issue>8</issue>
  <fpage>827</fpage>
  <lpage>38</lpage>
</bibl>

<bibl id="B12">
  <title><p>HTSeq: Analysis of high-throughput sequencing data with
  Python.</p></title>
  <aug>
    <au><snm>Anders</snm><fnm>S</fnm></au>
  </aug>
  <pubdate>2011</pubdate>
  <url>http://www-huber.embl.de/users/anders/HTSeq/</url>
</bibl>

<bibl id="B13">
  <title><p>Picard</p></title>
  <aug>
    <au><snm>Wysoker</snm><fnm>A</fnm></au>
    <au><snm>Tibbetts</snm><fnm>K</fnm></au>
    <au><snm>Fennell</snm><fnm>T</fnm></au>
  </aug>
  <pubdate>2012</pubdate>
  <url>http://picard.sourceforge.net/</url>
</bibl>

<bibl id="B14">
  <title><p>BEDTools: a flexible suite of utilities for comparing genomic
  features</p></title>
  <aug>
    <au><snm>Quinlan</snm><fnm>AR</fnm></au>
    <au><snm>Hall</snm><fnm>IM</fnm></au>
  </aug>
  <source>Bioinformatics</source>
  <pubdate>2010</pubdate>
  <volume>26</volume>
  <issue>6</issue>
  <fpage>841</fpage>
  <lpage>2</lpage>
</bibl>

<bibl id="B15">
  <title><p>A comprehensive evaluation of normalization methods for Illumina
  high-throughput RNA sequencing data analysis</p></title>
  <aug>
    <au><snm>Dillies</snm><fnm>MA</fnm></au>
    <au><snm>Rau</snm><fnm>A</fnm></au>
    <au><snm>Aubert</snm><fnm>J</fnm></au>
    <au><snm>Hennequet Antier</snm><fnm>C</fnm></au>
    <au><snm>Jeanmougin</snm><fnm>M</fnm></au>
    <au><snm>Servant</snm><fnm>N</fnm></au>
    <au><snm>Keime</snm><fnm>C</fnm></au>
    <au><snm>Marot</snm><fnm>G</fnm></au>
    <au><snm>Castel</snm><fnm>D</fnm></au>
    <au><snm>Estelle</snm><fnm>J</fnm></au>
    <au><snm>Guernec</snm><fnm>G</fnm></au>
    <au><snm>Jagla</snm><fnm>B</fnm></au>
    <au><snm>Jouneau</snm><fnm>L</fnm></au>
    <au><snm>Lalo{\"e}</snm><fnm>D</fnm></au>
    <au><snm>Le Gall</snm><fnm>C</fnm></au>
    <au><snm>Scha{\"e}ffer</snm><fnm>B</fnm></au>
    <au><snm>Le Crom</snm><fnm>S</fnm></au>
    <au><snm>Guedj</snm><fnm>M</fnm></au>
    <au><snm>Jaffr{\'e}zic</snm><fnm>F</fnm></au>
    <au><cnm>{on behalf of The French StatOmique Consortium}</cnm></au>
  </aug>
  <source>Brief Bioinform</source>
  <pubdate>2012</pubdate>
</bibl>

<bibl id="B16">
  <title><p>Evaluation of statistical methods for normalization and
  differential expression in mRNA-Seq experiments</p></title>
  <aug>
    <au><snm>Bullard</snm><fnm>JH</fnm></au>
    <au><snm>Purdom</snm><fnm>E</fnm></au>
    <au><snm>Hansen</snm><fnm>KD</fnm></au>
    <au><snm>Dudoit</snm><fnm>S</fnm></au>
  </aug>
  <source>BMC Bioinformatics</source>
  <pubdate>2010</pubdate>
  <volume>11</volume>
  <fpage>94</fpage>
</bibl>

<bibl id="B17">
  <title><p>A scaling normalization method for differential expression analysis
  of RNA-seq data</p></title>
  <aug>
    <au><snm>Robinson</snm><fnm>MD</fnm></au>
    <au><snm>Oshlack</snm><fnm>A</fnm></au>
  </aug>
  <source>Genome Biol</source>
  <pubdate>2010</pubdate>
  <volume>11</volume>
  <issue>3</issue>
  <fpage>R25</fpage>
</bibl>

<bibl id="B18">
  <title><p>A comparison of normalization methods for high density
  oligonucleotide array data based on variance and bias</p></title>
  <aug>
    <au><snm>Bolstad</snm><fnm>B M</fnm></au>
    <au><snm>Irizarry</snm><fnm>R A</fnm></au>
    <au><snm>Astrand</snm><fnm>M</fnm></au>
    <au><snm>Speed</snm><fnm>T P</fnm></au>
  </aug>
  <source>Bioinformatics</source>
  <pubdate>2003</pubdate>
  <volume>19</volume>
  <issue>2</issue>
  <fpage>185</fpage>
  <lpage>93</lpage>
</bibl>

<bibl id="B19">
  <title><p>Moderated statistical tests for assessing differences in tag
  abundance</p></title>
  <aug>
    <au><snm>Robinson</snm><fnm>MD</fnm></au>
    <au><snm>Smyth</snm><fnm>GK</fnm></au>
  </aug>
  <source>Bioinformatics</source>
  <pubdate>2007</pubdate>
  <volume>23</volume>
  <issue>21</issue>
  <fpage>2881</fpage>
  <lpage>7</lpage>
</bibl>

<bibl id="B20">
  <title><p>The transcriptional landscape of the yeast genome defined by RNA
  sequencing</p></title>
  <aug>
    <au><snm>Nagalakshmi</snm><fnm>U</fnm></au>
    <au><snm>Wang</snm><fnm>Z</fnm></au>
    <au><snm>Waern</snm><fnm>K</fnm></au>
    <au><snm>Shou</snm><fnm>C</fnm></au>
    <au><snm>Raha</snm><fnm>D</fnm></au>
    <au><snm>Gerstein</snm><fnm>M</fnm></au>
    <au><snm>Snyder</snm><fnm>M</fnm></au>
  </aug>
  <source>Science</source>
  <pubdate>2008</pubdate>
  <volume>320</volume>
  <issue>5881</issue>
  <fpage>1344</fpage>
  <lpage>9</lpage>
</bibl>

<bibl id="B21">
  <title><p>RNA-seq: an assessment of technical reproducibility and comparison
  with gene expression arrays</p></title>
  <aug>
    <au><snm>Marioni</snm><fnm>JC</fnm></au>
    <au><snm>Mason</snm><fnm>CE</fnm></au>
    <au><snm>Mane</snm><fnm>SM</fnm></au>
    <au><snm>Stephens</snm><fnm>M</fnm></au>
    <au><snm>Gilad</snm><fnm>Y</fnm></au>
  </aug>
  <source>Genome Res</source>
  <pubdate>2008</pubdate>
  <volume>18</volume>
  <issue>9</issue>
  <fpage>1509</fpage>
  <lpage>17</lpage>
</bibl>

<bibl id="B22">
  <title><p>TopHat: discovering splice junctions with RNA-Seq</p></title>
  <aug>
    <au><snm>Trapnell</snm><fnm>C</fnm></au>
    <au><snm>Pachter</snm><fnm>L</fnm></au>
    <au><snm>Salzberg</snm><fnm>SL</fnm></au>
  </aug>
  <source>Bioinformatics</source>
  <pubdate>2009</pubdate>
  <volume>25</volume>
  <issue>9</issue>
  <fpage>1105</fpage>
  <lpage>11</lpage>
</bibl>

<bibl id="B23">
  <title><p>Evaluation of DNA microarray results with quantitative gene
  expression platforms</p></title>
  <aug>
    <au><snm>Canales</snm><fnm>RD</fnm></au>
    <au><snm>Luo</snm><fnm>Y</fnm></au>
    <au><snm>Willey</snm><fnm>JC</fnm></au>
    <au><snm>Austermiller</snm><fnm>B</fnm></au>
    <au><snm>Barbacioru</snm><fnm>CC</fnm></au>
    <au><snm>Boysen</snm><fnm>C</fnm></au>
    <au><snm>Hunkapiller</snm><fnm>K</fnm></au>
    <au><snm>Jensen</snm><fnm>RV</fnm></au>
    <au><snm>Knight</snm><fnm>CR</fnm></au>
    <au><snm>Lee</snm><fnm>KY</fnm></au>
    <au><snm>Ma</snm><fnm>Y</fnm></au>
    <au><snm>Maqsodi</snm><fnm>B</fnm></au>
    <au><snm>Papallo</snm><fnm>A</fnm></au>
    <au><snm>Peters</snm><fnm>EH</fnm></au>
    <au><snm>Poulter</snm><fnm>K</fnm></au>
    <au><snm>Ruppel</snm><fnm>PL</fnm></au>
    <au><snm>Samaha</snm><fnm>RR</fnm></au>
    <au><snm>Shi</snm><fnm>L</fnm></au>
    <au><snm>Yang</snm><fnm>W</fnm></au>
    <au><snm>Zhang</snm><fnm>L</fnm></au>
    <au><snm>Goodsaid</snm><fnm>FM</fnm></au>
  </aug>
  <source>Nat Biotechnol</source>
  <pubdate>2006</pubdate>
  <volume>24</volume>
  <issue>9</issue>
  <fpage>1115</fpage>
  <lpage>22</lpage>
</bibl>

<bibl id="B24">
  <title><p>Detecting differential usage of exons from RNA-seq data</p></title>
  <aug>
    <au><snm>Anders</snm><fnm>S</fnm></au>
    <au><snm>Reyes</snm><fnm>A</fnm></au>
    <au><snm>Huber</snm><fnm>W</fnm></au>
  </aug>
  <source>Genome Res</source>
  <pubdate>2012</pubdate>
  <volume>22</volume>
  <issue>10</issue>
  <fpage>2008</fpage>
  <lpage>17</lpage>
</bibl>

<bibl id="B25">
  <title><p>Efficient experimental design and analysis strategies for the
  detection of differential expression using RNA-Sequencing</p></title>
  <aug>
    <au><snm>Robles</snm><fnm>JA</fnm></au>
    <au><snm>Qureshi</snm><fnm>SE</fnm></au>
    <au><snm>Stephen</snm><fnm>SJ</fnm></au>
    <au><snm>Wilson</snm><fnm>SR</fnm></au>
    <au><snm>Burden</snm><fnm>CJ</fnm></au>
    <au><snm>Taylor</snm><fnm>JM</fnm></au>
  </aug>
  <source>BMC Genomics</source>
  <pubdate>2012</pubdate>
  <volume>13</volume>
  <issue>1</issue>
  <fpage>484</fpage>
</bibl>

<bibl id="B26">
  <title><p>A comparison of statistical methods for detecting differentially
  expressed genes from RNA-seq data</p></title>
  <aug>
    <au><snm>Kvam</snm><fnm>VM</fnm></au>
    <au><snm>Liu</snm><fnm>P</fnm></au>
    <au><snm>Si</snm><fnm>Y</fnm></au>
  </aug>
  <source>Am J Bot</source>
  <pubdate>2012</pubdate>
  <volume>99</volume>
  <issue>2</issue>
  <fpage>248</fpage>
  <lpage>56</lpage>
</bibl>

<bibl id="B27">
  <title><p>The birth of the Epitranscriptome: deciphering the function of RNA
  modifications</p></title>
  <aug>
    <au><snm>Saletore</snm><fnm>Y</fnm></au>
    <au><snm>Meyer</snm><fnm>K</fnm></au>
    <au><snm>Korlach</snm><fnm>J</fnm></au>
    <au><snm>Vilfan</snm><fnm>ID</fnm></au>
    <au><snm>Jaffrey</snm><fnm>S</fnm></au>
    <au><snm>Mason</snm><fnm>CE</fnm></au>
  </aug>
  <source>Genome Biol</source>
  <pubdate>2012</pubdate>
  <volume>13</volume>
  <issue>10</issue>
  <fpage>175</fpage>
</bibl>

</refgrp>
} 

\subsection*{Acknowledgments}
D.B. is supported by grants from the Starr and DeGregorio Family foundations. F.R., R.K., Y.L., A.K., N.D.S. were supported by MSKCC Comprehensive Cancer Center (P30 CA008748) and by the director of the Sloan-Kettering Institute. Additionally F.R. is supported by the Susan and Peter Solomon Divisional Genomics Program. R.K. and N.D.S are supported by the MSKCC SPORE in Prostate Cancer (P50 CA091629), R.K. is supported by the Clinical Translational Science Center and N.D.S is supported by the SPORE in Soft Tissue Sarcoma (P50 CA140146). The authors greatly acknowledge Weill Cornell Epigenomics Core contribution.

\end{document}